\documentclass[aps,prd,superscriptaddress,12pt,showpacs,notitlepage]{revtex4}
	\usepackage{amsmath,amssymb,mathrsfs}
	\usepackage[dvips]{graphicx}
	\usepackage{subfig}
	
\usepackage{epsf}
\usepackage{epsfig}

\newcommand{\be}{\begin{equation}}
\newcommand{\ee}{\end{equation}}
\newcommand{\bea}{\begin{eqnarray}}
\newcommand{\eea}{\end{eqnarray}}
\newcommand{\pa}{\partial}
\newcommand{\bb}{\bibitem}
 
\newcommand{\eqn}{\begin{eqnarray}}
\newcommand{\eqnx}{\end{eqnarray}}

\begin{document}
\title{Lifshitz field theories with SDiff symmetries}
\author{C. Adam}
\affiliation{Departamento de F\'isica de Part\'iculas, Universidad de Santiago de Compostela and Instituto Galego de F\'isica de Altas Enerxias (IGFAE) E-15782 Santiago de Compostela, Spain}
\author{C. Naya}
\affiliation{Departamento de F\'isica de Part\'iculas, Universidad de Santiago de Compostela and Instituto Galego de F\'isica de Altas Enerxias (IGFAE) E-15782 Santiago de Compostela, Spain}
\author{J. Sanchez-Guillen}
\affiliation{Departamento de F\'isica de Part\'iculas, Universidad de Santiago de Compostela and Instituto Galego de F\'isica de Altas Enerxias (IGFAE) E-15782 Santiago de Compostela, Spain}
\author{A. Wereszczynski}
\affiliation{Institute of Physics,  Jagiellonian University,
Reymonta 4, Krak\'{o}w, Poland}

\pacs{11.30.Pb, 11.27.+d}

\begin{abstract}
We consider Lifshitz field theories with a dynamical critical exponent $z$ equal to the dimension of space $d$ and with a large group of base space symmetries, concretely space coordinate transformations with unit determinant ("Special Diffeomorphisms"). The field configurations of the theories considered may have the topology of skyrmions, vortices or monopoles, although we focus our detailed investigations on skyrmions. The resulting Lifshitz field theories have a BPS bound and exact soliton solutions saturating the bound, as well as time-dependent topological Q-ball solutions. Finally, we investigate the U(1) gauged versions of the Lifshitz field theories coupled to a Chern--Simons gauge field, where the BPS bound and soliton solutions saturating the bound continue to exist.  
\end{abstract}

\maketitle 
%%%%%%%%%%%%%%%%%%%%%%%%%%%%%%%%%%%%%%%%%
\section{Introduction}
%%%%%%%%%%%%%%%%%%%%%%%%%%%%%%%%%%%%%%%%%
The last years have seen a rising interest in applying the original  Lifshitz  \cite{L} concept of anisotropic scaling of condensed matter physics \cite{H} in other areas of theoretical physics from field theory \cite {FT} to gravitation, where it opened a new approach to quantum gravity \cite {G}, and string theory (ADS/CFT) \cite{horava3}, and back to modern condensed matter theory \cite{Tong}. The idea offers an explicit simple realization of systems having different symmetries and energy behaviours at short and long distances. Specifically, for the case of quantum field theories, an anisotropic scaling of space and time offers the possibility to improve  perturbative renormalizability.
These intense activities have achieved  many results like a new control of Lorentz violations,  IR limits of UV renormalizable gravity theories or interesting ABJM Chern-Simons duals, but also stumbled on problems, like detailed balance, stability of black holes, difficulties with symmetric astrophysical solutions, or the persistence of unwanted degrees of freedom \cite{blas}. These can be circumvented for the moment with special readjustments and probably are a reflection of general deep problems \cite {Soti}. Therefore, a better understanding of  the basic field theory input, preferably with exact results even with simplified models, may be useful.
There have been in fact recent works \cite {Plata} dealing with solvable (numerically for the moment) $3d$ Lifshitz models with scalars, with gauge couplings  or including a Chern-Simon term to avoid higher order terms.

In this spirit we present  here exact solutions and BPS  stability discussions of specific Lifshitz field theories (BPS solitons in Lifshitz theories have also been found in \cite{kobak}), related naturally to special 
solvable topological soliton models without  quadratic terms (quenched) \cite{BPS-Sk} - \cite{top-dual}, with special diffeomorphism (SDiff) invariance, which, as we will show, contain in a way  some of the anisotropic scaling features, like symmetries. Here the emphasis is on nonperturbative  solutions carrying a topological charge, which may be of the skyrmion or vortex type, and which may be compact or noncompact, depending on the potentials considered and their vacuum structure. It is interesting that one obtains exact solutions in the specific Lifshitz models considered here. In fact, any exact result in this context can be of interest. 

The paper is organized as follows. In Section II we discuss general scaling properties of Lifshitz field theories and explain how the topological terms which typically exist in soliton models may be naturally incorporated into Lifshitz field theories with dynamical critical exponent $z$ equal to the space dimension $d$. We further show that these Lifshitz soliton models contain submodels with invariance under the infinite-dimensional group of area or volume preserving diffeomorphisms (SDiff symmetry, in general). In Section III, we study in detail a Lifshitz baby Skyrme model which has a BPS bound and both static soliton and time-dependent topological Q-ball solutions. In Section IV we study the gauged version of this Lifshitz baby Skyrme model with a Chern-Simons interaction of the gauge field. The model, again, has a BPS bound and soliton solutions saturating the bound, and this BPS bound requires the introduction of a superpotential like in supergravity \cite{susy1}, \cite{f susy}, analogously to the case of the gauged BPS baby Skyrme model \cite{gaugedBPSbaby}. Finally, Section V contains our conclusions.
%%%%%%%%%%%%%%%%%%%%%%%%%%%%%%%%%%%%%%%
\section{Lifshitz scaling and SDiff symmetries}
%%%%%%%%%%%%%%%%%%%%%%%%%%%%%%%%%%%%%%%
It is well-known that the presence of higher derivatives in the terms of a field theory lagrangian which contribute to the propagator (i.e., terms quadratic in the fields and their derivatives) improves the UV behaviour of the theory, because the propagator in momentum space will contain higher powers of momenta in the denominator. A Lorentz-invariant implementation of this idea, however, faces the problem that higher than first time derivatives in the lagrangian in general spoil unitarity. This lead to the recent proposal \cite{FT}, \cite{G} of field theories with an anisotropic scaling between space and time coordinates, sacrificing thereby (at least in the deep UV region) Lorentz invariance in favor of renormalizability and unitarity. Let us briefly discuss the simple example of a real scalar field in $d+1$ space-time dimensions. For the moment, we consider the action consisting of the following two terms
\be \label{action-one}
S= \frac{1}{2}\int d^d x dt \left( {\dot \Phi}^2 - \Phi (-\Delta)^k \Phi \right)
\ee
where $\dot\phi$ is the time derivative of $\phi$, $\Delta$ is the laplacian and $k$ is a positive integer. For $k=1$, we recover the standard kinetic term for a scalar field upon a partial integration, but here we shall consider the case $k>1$.  The action may (and, in general, will) contain more terms, but the two terms above are special in that they determine the scaling symmetry and the perturbative renormalizability properties of the theory. Concretely, let us discuss the scaling behaviour. If we introduce the scaling transformations
\be
t \to \kappa^{-z} t , \quad \vec x \to \kappa^{-1} \vec x , \quad \Phi \to \kappa^{\varphi} \Phi
\ee
and require invariance of both terms in the action, then we easily find the following relations,
\be
2 \varphi = d-k , \quad z=k.
\ee
The two terms $\dot\Phi^2$ and $\Phi (-\Delta)^k \Phi$ in the lagrangian density corresponding to the above action both have scaling dimension $d+k$. We may add further terms to the lagrangian density, but they should not exceed this scaling dimension. Concretely, terms with scaling dimension $d+k$ correspond to renormalizable, scale invariant interactions, whereas terms with a smaller scaling dimension correspond to super-renormalizable interactions. 

In a next step, let us make some further assumptions which are adequate for our purposes. First of all, we assume from now on that $k=d$. This has the consequence that the scalar field is now dimensionless, $\varphi =0$. We may, therefore, add arbitrary potential terms $V(\Phi)$ to the lagrangian density without spoiling renormalizability. We may also multiply derivative terms by functions of $\Phi$ like, e.g., $f(\Phi) (-\Delta)^{k'} \Phi$ where, obviously, $k' \le k$ must hold. But this last observation opens the possibility to include nonlinear terms into the interaction which typically show up in field theories with topological solitons like, e.g., the Skyrme model and related theories.  We simply have to consider more than one scalar field and to act with derivatives on them as required by these Skyrme terms. Concretely, we want to consider terms which are related to topological charge or winding number densities, which in $d$ space dimensions have the typical form
\be
{\cal Q}_f = f(\Phi^a) \epsilon^{j_1 \ldots j_d}\epsilon^{a_1 \ldots a_d} \pa_{j_1}\Phi^{a_1} \ldots \pa_{j_d} \Phi^{a_d}.
\ee
Here, $f(\Phi^a)$ is a function of the fields which depends on the geometry and topology of the target space manifold. Further, we assumed that the number of fields $\Phi^a$ is equal to the space dimension $d$, $a=1 , \ldots ,d$, which is the most relevant case for topological solitons. The square of ${\cal Q}_f$ is obviously positive semi-definite. In addition, ${\cal Q}_f^2$ has scaling dimension $2d$ (where we assume $k=d$) and corresponds, therefore, to a renormalizable, scale invariant interaction.   

Up to now, we mainly discussed the effect which Lifshitz-type field theories with anisotropic scaling have on perturbative renormalizability, but now we shall slightly change our point of view. That is to say, we want to consider Lifshitz field theories not just as a fix for renormalization problems, but instead as proper field theories in their own right, which may display new symmetries or new nonperturbative features not present in standard, Lorentz-invariant field theories. If the proposal of a field theory with anisotropic scaling in the UV is to be taken seriously, then such an investigation is certainly required.  Concretely, we want to investigate the possible existence of BPS bounds and soliton and Q-ball solutions for a class of Lifshitz-type submodels which are characterized by a large group of symmetries. The action for the submodels has the generic form
\be \label{sub-lifshitz-action}
S=\frac{1}{2} \int d^d x dt \left( g_{ab}( \Phi^c) {\dot \Phi}^a {\dot \Phi }^b -\lambda^2 {\cal Q}_f^2 - \mu^2 V(\Phi^a) \right) .
\ee
Here, $\lambda$ and $\mu$ are real coupling constants, and $g_{ab}$ is a matrix of functions of $\Phi^a$ ("target space metric") which obeys $\det g_{ab}> 0$. The two terms $g_{ab}( \Phi^c) {\dot \Phi}^a {\dot \Phi }^b$ and ${\cal Q}_f^2$ have scaling dimensions $2d$, whereas the potential $V(\Phi^a)$ is dimensionless. The absence of a term of the type $ \Phi (-\Delta)^k \Phi  $ implies that these submodels are certainly not useful for the discussion of perturbative features, but they may be useful to understand some nonperturbative properties of Lifshitz-type theories. The reason to choose the above actions is that they have an infinite-dimensional group of symmetries. They are invariant under base space SDiff ("Special Diffeomorphism") transformations, i.e., diffeomorphisms of the space coordinates $x^j \to x'^j = y^j (x^k)$ which leave the volume form $d^d x$ invariant (i.e., diffeomorphisms with unit determinant). We remark that in Horava-Lifshitz gravity \cite{G} the action is invariant under foliation-preserving diffeomorphisms, therefore our above actions may be seen as toy models for the study of certain nonperturbative issues of that proposal after a partial gauge fixing (setting the lapse function equal to one and the shift functions equal to zero, see also \cite{bakas1}, \cite{bakas2} for further discussion).  The infinitely many symmetries (and corresponding conservation laws) of the above actions (\ref{sub-lifshitz-action}) imply that these theories are integrable in the sense of generalized integrability \cite{gen-int}, and, therefore, one expects the existence of exact nonperturbative solutions.  

If we restrict to static field configurations, then the corresponding energy functional 
\be \label{sub-lifshitz-energy}
E[\phi^a ]=\frac{1}{2} \int d^d x dt \left( \lambda^2 {\cal Q}_f^2 + \mu^2 V(\Phi^a) \right) 
\ee
has even more symmetries. Indeed, the first term is invariant under the group of target space SDiff symmetries, that is, under the target space diffeomorphisms which leave the target space volume form $\Omega = f(\Phi^a) d\Phi^1 \ldots d\Phi^d$ invariant. The potential, in general, breaks some of these SDiff symmetries, but for sufficiently symmetric potentials an infinite-dimensional symmetry group remains. For potentials, e.g., which only depend on the modulus $\Phi^a \Phi^a$, all SDiff transformation which do not change this modulus remain symmetries of the full static energy.   It is interesting to compare the symmetries of the actions (\ref{sub-lifshitz-action}) with the symmetries of the BPS-Skyrme type theories \cite{BPS-Sk} - \cite{top-dual} whose actions are simply the Lorentz-invariant generalizations to time-dependent field configurations of the above energy (\ref{sub-lifshitz-energy}), i.e., their actions have the form
 \be \label{BPS-skyrme-action}
S=\frac{1}{2} \int d^d x dt \left( -\lambda^2 ({\cal Q}_f^\mu)^2 - \mu^2 V(\Phi^a) \right) 
\ee
where
\be
{\cal Q}_f^\mu = f(\Phi^a) \epsilon^{\mu \mu_1 \ldots \mu_d}\epsilon^{a_1 \ldots a_d} \pa_{\mu_1}\Phi^{a_1} \ldots \pa_{\mu_d} \Phi^{a_d}
\ee
is a Lorentz-covariant topological current.
The static energy functionals of the two types of theories are the same, therefore also their symmetries coincide. For the time dependent situation, on the other hand, the roles of the two groups of symmetries are inverted for the two types of theories. That is to say, while the actions (\ref{sub-lifshitz-action}) still have the base space SDiffs as symmetries (these are, therefore, Noether symmetries), but no longer the target space SDiffs (more precisely, the subgroup which leaves the potential invariant), the situation is exactly the other way round for the theories (\ref{BPS-skyrme-action}). These actions are still invariant under the target space SDiff subgroup, but no longer under the base space SDiffs. Another difference between the Lifshitz-type theories (\ref{sub-lifshitz-action}) and the BPS-Skyrme type theories (\ref{BPS-skyrme-action}) is that the former have a well-defined Cauchy problem, whereas the latter ones don't (i.e., for some initial data their time evolution is not well-defined, see e.g. \cite{fosco}, \cite{top-dual}). 

Finally, we shall restrict to the case of two space dimensions, $d=2$, for the concrete examples considered below. The main reason is that calculations are simpler there, although for most results the generalization to higher dimensions should pose no problem. Besides, the field configurations in the examples considered will have the topology of (baby) skyrmions. We could also easily generalize to fields with the topology of vortices (or monopoles in $d=3$ dimensions) - indeed, for the BPS Skyrme type models (\ref{BPS-skyrme-action}) there even exists an exact topological duality between theories and solutions with Skyrme (soliton) topology and theories and solutions with vortex (or monopole) topology \cite{top-dual}. 
But for reasons of simplicity we restrict our examples to baby skyrmions.

%%%%%%%%%%%%%%%%%%%%%%%%%%%%%%%%%%%%%%%%%
\section{The Lifshitz BPS baby Skyrme model}
%%%%%%%%%%%%%%%%%%%%%%%%%%%%%%%%%%%%%%%%%
For baby skyrmions \cite{baby1}, the target space manifold is the unit two-sphere $S^2$. This implies that the fields $\Phi^a$, $a=1,2$ should be interpreted as the real and imaginary part, $u=\Phi^1 + i \Phi^2$, of a stereographic projection (or CP(1)) coordinate $u\in \mathbb{C}$. The target space metric and area two-form are, therefore,
\be
g_{ab} = \frac{4}{(1+(\Phi^1)^2 + (\Phi^2)^2)^{2}} \delta_{ab} ,\quad d\Omega =  \frac{4}{(1+(\Phi^1)^2 + (\Phi^2)^2)^{2}} d\Phi^1 d\Phi^2 
\ee
and the topological current is
\be
{\cal Q}_f ^\mu \longrightarrow q^\mu = \frac{2}{(1+(\Phi^1)^2 + (\Phi^2)^2)^{2}} \epsilon^{\mu\nu\lambda}\epsilon^{ab}\pa_\nu \Phi^a \pa_\lambda \Phi^b
\ee
and provides the degree of the map defined by the Skyrme field $\Phi^a (\vec x)$ (topological charge or winding number) according to
\begin{equation}
\mbox{deg} [\vec{\phi}]=\frac{1}{4\pi} \int d^2x q \; \in \; Z
\end{equation}
where we use the short-hand notation $q\equiv q^0$. In the following, however, we prefer to parametrize the target space two-sphere by a three-component unit vector $\vec \phi$, i.e., $\vec \phi^2=1$ related to $u=\Phi^1 + i \Phi^2$ by stereographic projection. In this parametrization, there are no nontrivial target space metric and area factors, and the nontrivial target space geometry is taken into account by the constraint $\vec \phi^2 =1$. E.g., the topological current reads
\begin{equation}
\epsilon_{\mu\nu\lambda}q^\lambda \equiv
q_{\mu\nu}=\vec{\phi}\cdot \left(\partial_\mu \vec{\phi} \times \partial_\nu \vec{\phi} \right) .
\end{equation}  

%%%%%%%%%%%%%%%%%%%%%%%%%%%%
\subsection{The model}
%%%%%%%%%%%%%%%%%%%%%%%%%%%%
In contrast to the BPS baby Skyrme model \cite{GP} - \cite{Sp1}, its Lifshitz version has the standard time derivative part in the Lagrangian (but remember the constraint $\vec \phi^2 =1$),  while the term containing the space derivatives remains unchanged,
\begin{equation}
\mathcal{L}= \gamma (\partial_0 \vec{\phi})^2-\frac{\lambda^2}{2} q^2 - \mu^2 V(\vec{n}\cdot \vec{\phi} ) .
\end{equation}
Here, the potential $V(\vec n \cdot \vec \phi)$ breaks the SO(3) target space symmetry (rotations of $\vec \phi$) down to an SO(2) symmetry (rotations about the $\vec n$ axis). Further, the potential obeys $V\ge 0$, $V(1)=0$, and $\vec n$ is the vacuum vector which $\vec \phi$ must approach in the large distance limit for finite energy field configurations, $\lim_{|\vec x| \to \infty} \vec \phi = \vec n$. Further, $\gamma$, $\lambda$ and $\mu$ are real constants.

In the static limit, the model is identical to the BPS baby Skyrme model and, therefore, possesses the same exact static solutions, which, depending on the potential $V$, are compactons or usual infinitely extended solitons with power-like or exponential localization. Let us briefly summarize some known results \cite{BPS baby}. For the family of old baby potentials (one-vacuum potentials) 
 \begin{equation}
 V=\left(\frac{1-\phi^3}{2} \right)^a
 \end{equation}
 where $a \in (0,\infty)$, we get compactons ($a\in (0,2)$), one exponential solution ($a=2$) and power-like localized solitons ($a >2$). We remark that the effect of equivalent potentials for generalizations with a CP$^N$ target space have been studied recently in \cite{kli-saw}. On the other hand, for the family of the new baby potentials (two-vacuum potentials)    
  \begin{equation}
 V=\left(1-(\phi^3)^2 \right)^a
 \end{equation}
 one finds only compactons for $a \in (0,2)$. For other potentials \cite{baby pot}, \cite{comp-bS} and exact solutions see \cite{BPS baby}.
 \\
 While the static spectra of the BPS baby Skyrme model and the Lifshitz version coincide, their time dependent solutions as well as the dynamical properties of the static configurations differ dramatically. In the case of the BPS baby Skyrme model, the time evolution is a highly non-trivial problem since the equations are not globally hyperbolic. Hence, for some initial configurations time evolution may be not globally well defined and could lead to singularities. However, due to the fact that the Lagrangian depends only on first time derivatives squared one can still define a proper Hamiltonian and (which is important from the point of view of the application to the nuclear physics) perform the semiclassical quantization. Obviously, the Lifshitz version has a well defined dynamics which can be at least treated using numerical methods.  
 %%%%%%%%%%%%%%%%%%%%%%%%%%%%
\subsection{Dynamics}
%%%%%%%%%%%%%%%%%%%%%%%%%%%%
The full time dependent equations of motion read
\begin{equation}
 -2\gamma \vec{\phi} \times \partial_0^2 \vec{\phi} + \mu^2 \vec{n}
\times \vec{\phi} V' + \lambda^2 \nabla_i \vec K_i=0
\end{equation}
where
\begin{equation}
\vec K_i= \nabla_j \vec{\phi}  \left( \vec{\phi} \cdot (\nabla_i \vec{\phi} \times \nabla_j \vec{\phi}) \right) .
\end{equation}
Let us observe that the Lagrangian looks like the classical membrane Lagrangian with a nontrivial potential \cite{fairlie}
\begin{equation}
\mathcal{L}= \gamma (\partial_0 X^a)^2-\frac{\lambda^2}{4} \{ X^a,X^b\}^2 - \mu^2 V(X^a ) 
\end{equation}
where
\begin{equation}
\{ X^a,X^b\} = \epsilon_{ij} \partial_iX^a \partial_j X^b
\end{equation}
is the Poisson bracket on the base space while $X^a\equiv \phi^a$. Then, the static Bogomolny equation is just  
\begin{equation}
\epsilon^{abc} \{ X^a,X^b\}  \pm \sqrt{V} X^a=0
\end{equation}
and is identical to the BPS equation of the BPS baby Skyrme model.
On the other hand, for the usual membrane (with potential being in fact a constraint - see, e.g., $V=(X^a)^2$), the second order dynamics can be reduced to a first order Nahm equation
\begin{equation}
\partial_0 X^a=\epsilon^{abc} \{ X^a,X^b\}  + X^a .
\end{equation}
Due to the similarity with our Bogomolny equation, one might hope that also here the dynamics could be integrated to a first order "nonlinear" Nahm equation
\begin{equation}
\partial_0 X^a=\epsilon^{abc} \{ X^b,X^c\}  + \sqrt{V}X^a . \label{non Nahm}
\end{equation}
This is, however, not true, and the dynamics of the two theories is completely different. The easiest way to see it is to multiply the last formula by $X^a$, i.e.,
\begin{equation}
X^a \partial_0 X^a= \epsilon^{abc} X^a \{ X^b,X^c\}  + \sqrt{V}.
\end{equation}
If we now assume that the $X^a$ span a two-sphere, then $X^a \dot X^a =0$, the r.h.s. is exactly the Bogomolny equation and, as $ X^d (\epsilon^{abc} X^a \{ X^b,X^c\} ) = \epsilon^{dbc} \{ X^b,X^c\}$ we find that 
\begin{equation}
\epsilon^{abc} \{ X^a,X^b\}  + \sqrt{V}X^a=0 
\end{equation}
and, hence
\begin{equation}
\partial_0 X^a=0
\end{equation}
i.e., no dynamics in the system. This just corresponds to the well-known fact that the sphere is a vacuum solution of the membrane action.
On the other hand, we expect nontrivial dynamics for our Lifshitz model with the target space manifold $S^2$, so the dynamics (time dependence) of the two theories is certainly different.

%%%%%%%%%%%%%%%%%%%%%%%%%%%%
\subsection{Spinning compacton solutions}
%%%%%%%%%%%%%%%%%%%%%%%%%%%%
It is possible to find exact time dependent solutions assuming the following axially symmetric ansatz ($x^1 = r\cos \phi$, $x^2 = r\sin \phi$)
\begin{equation} 
\vec{\phi} (r,\phi,t)  = \left( 
\begin{array}{c}
\sin f(r) \cos (n\phi - \omega t) \\
\sin f(r) \sin ( n\phi - \omega t) \\
\cos f(r)
\end{array}
\right)
\end{equation}
where $\omega$ is a real parameter of the internal rotation. Here, the integer $n$ is equal to the topological charge of the skyrmion provided that the profile function $f(r)$ satisfied the corresponding skyrmion boundary conditions. 
The resulting Lagrangian reads
\begin{equation}
L=2\pi \int dx \left( 4\gamma \omega^2 h(1-h)-2 \lambda^2 n^2 h_x^2  -\mu^2 V(h) \right)
\end{equation}
where we have assumed $\vec n = (0,0,1) \; \Rightarrow \; \vec n \cdot \vec \phi = \phi_3$ and used the new target space variable
\begin{equation} \label{h-def}
h= \frac{1}{2}(1-\phi_3)
\end{equation}
Moreover, $x=r^2/2$. The pertinent equation of motion is
\begin{equation}
-4\lambda^2 n^2 h_{xx}+\mu^2V_h - 4\gamma \omega^2 (1-2h)=0
\end{equation}
where the nontrivial topology of the Skyrme field enforces the following boundary conditions
\begin{equation}
h(x=0)=1, \;\; h(x=X)=0, \;\;\; h_x(x=X)=0
\end{equation}
where $X$ can be finite (compactons) or infinite (usual soliton).
Taking into account the boundary conditions, this equation can be easily integrated to
\begin{equation}
2\lambda^2 n^2 h_{x}^2=\mu^2V - 4\gamma \omega^2 h(1-h) . \label{h static}
\end{equation}
Of course, one can integrate this formula again and easily express the solution in terms of the 
integral depending on the potential $V$. Hence, similarly to the static case we find exact solutions.
\\ 
It should be stressed that in order to have a topologically nontrivial solution $h$ must cover the segment $[0,1]$ which results in a condition for the effective potential. Namely, 
\begin{equation}
V_{eff}=\frac{\mu^2}{n^2}V - 4\frac{\gamma \omega^2}{n^2} h(1-h)
\end{equation}
must be positive definite on the whole segment $h \in [0,1]$. This can restrict the possible values of $\omega$ as well as exclude the existence of the spinning solutions completely.  
\\  \\
{\bf Proposition 1:}  Generic spinning solutions are of a compact type while the critical solutions are compactons or infinitely extended solitons (exponentially or power-like localized).
\\
{\bf Proof:}
The approach to the vacuum, and therefore the qualitative type of a solitonic solution, is controlled by the asymptotical behavior of the effective potential near the vacuum. Here, in order to have finite energy configurations, we assume that $V$ (and in a consequence $V_{eff}$) must possesses one or two vacua located at $h=0$ (respectively at $h=0,1$). As the second term in the effective potential is fixed, it is enough to analyze the Taylor expansion of the original potential at $h=0$
\begin{equation}
V=c_0h^c+o(h^c).
\end{equation}
If $c<1$ then also the leading term of $V_{eff}$ goes sublinearly with the field. Due to that one obtains a compacton (which ceases to exist for sufficiently large $\omega$). Moreover, there are no spinning solutions for any potentials with $c>1$. In this case the effective potential has the linear leading linear term, however, with a negative prefactor. In other words, at $h=0$ the effective potential always gets a negative value. The last possibility when $c=1$ is slightly more subtle. Generically one finds again compactons as $V_{eff}$ is linear in $h$. However, there is a unique value of the frequency $\omega$ (we called it the critical frequency $\omega_{crit}$) for which the linear leading term in $V$ cancels with the linear term of the second part of $V_{eff}$
\begin{equation}
\omega_{crit}=\frac{\mu^2}{4\gamma} V'_h(0)
\end{equation}
Then, one has to take into account the next term in the expansion
\begin{equation}
V=c_0h+ c_1h^{d_1}+o(h^{d_1}).
\end{equation}
If $d_1<2$ then we get a compacton also in the critical limit ($c_1$ must be positive). For $d_1 > 2$ we find an exponentially localized solution as $(4\gamma \omega^2 /n^2)h^2$ becomes the leading term in $V_{eff}$. In the case $d_1=2$ there are two possibilities: if there are no other terms in the expansion we get a soliton with the exponential tail (under the condition that $c_1+4\gamma \omega_{crit}^2 /\mu^2>0$). If further higher terms in the expansion exist, power-like localized solutions are observed for $c_1 = -c_0$. For $\omega > \omega_{crit}$ no solutions exist. Obviously, such non-compact solutions are quite special (not generic) since they emerge for a unique value of the spinning frequency and/or for very special potentials.       
\\  \\
{\bf Observation 1:}  The old baby type potentials generating static BPS baby skyrmions with exponential or power-like localization do not lead to solitonic spinning solutions 
\\ \\
As an example with exact solution let us consider the old baby potential $V=h$. Then we find
\begin{equation}
h_{x}^2=\frac{1}{n^2} h(\alpha +\beta h)
\end{equation}
where 
\begin{equation}
\alpha^2 = \frac{1}{2\lambda^2} \left(\mu^2-4 \gamma \omega^2\right), \;\;\;\; \beta^2 =\frac{2\gamma \omega^2}{\lambda^2}
\end{equation}
In order to have a topologically nontrivial solution $h$ must cover the segment $[0,1]$ which requires $\alpha >0$. Hence we get an upper  bound for the frequency of the oscillation 
\begin{equation}
\omega \leq \omega_{crit} = \frac{\mu^2}{4\gamma}
\end{equation}
Then the exact compacton solution is given by
\begin{equation}
h= \left\{ 
\begin{array}{cc}
\frac{\alpha^2}{\beta^2} \sinh^2 \frac{\beta}{2n} (x-X) & x \leq X \\
0 & x \geq X
\end{array} \right.
\end{equation}
where the compacton boundary 
$$X=\frac{2n}{\beta} \mbox{arc sinh} \; \frac{\beta}{\alpha}$$
In the critical point where $\omega=\omega_{crit}$ and $\alpha=0$ we find a non-compacton exponentially localized solution
\begin{equation}
h_{crit}=e^{-\frac{\mu}{\sqrt{2} \lambda n }x}
\end{equation}
This is consistent with the fact that the size of the compacton grows to infinity as we approach the critical frequency. It is an interesting observation that the localization type of the solution depends not only on the original potential but also on the value of the oscillation frequency as both contribute to the effective  potential. 
\\
Similarly, for the new baby potential $V=h(1-h)$ we find 
\begin{equation}
h_{x}^2= \frac{1}{2\lambda^2n^2} \left( \mu^2-4\gamma \omega^2 \right) h (1-h)
\end{equation}
which represents a compacton with the same bound on the frequency. The corresponding compact baby Skyrmion reads
\begin{equation}
h= \left\{ 
\begin{array}{cc}
\sin^2 \frac{\alpha}{2n} (x-X) & x \leq X \\
0 & x \geq X
\end{array} \right.
\end{equation}
and
$$X=\frac{\pi n}{\alpha} $$
Using the first order profile equation we can rewrite the total energy integral of the spinning solutions as
\begin{equation}
E=2\pi \int dx \left( 4\gamma \omega^2 h(1-h)+2 \lambda^2n^2 h_x^2  +\mu^2 V(h) \right) = 4\pi \mu^2 \int dx V(h)
\end{equation}
which for the above considered old baby potential gives
$$E_{old} =  \frac{n}{\sqrt{\beta}} \left( \sqrt{\frac{\beta}{\alpha}} \sqrt{1+\frac{\beta}{\alpha}} -\mbox{arc sinh} \; \sqrt{\frac{\beta}{\alpha}} \right)$$
while the new baby potential leads to
$$ E_{new}=\frac{\pi^2 \mu^2 n }{2 \alpha} = \frac{\pi^2 \lambda \mu}{\sqrt{2}} \frac{n}{\sqrt{1-\frac{4\gamma \omega^2}{\mu^2}}}$$
Hence, both case lead to linear energy-topological charge relations, with the critical frequency fixed by the parameters of the model (and independent of the topological charge $Q=n$). This should be contrasted with the baby Skyrme model and its energies of spinning solutions \cite{comp-bS}, where the energy-charge relation is linear only approximately. In fact, $E/Q$ depends on the charge which leads to some instabilities for axially symmetric higher charge solutions. Moreover, there has been observed a symmetry breaking transition from the axially symmetrical baby skyrmions to discrete symmetry ones if the values of the model parameters are varied. The linear energy-charge relation found here may point toward the stability of the axially symmetric solutions. 
\\
As we see, the Lifshitz BPS baby Skyrme model is in a sense a non-trivial combination of the usual Skyrme model and its BPS version sharing some properties with former while others with the latter model. 
\\
In common with the BPS baby Skyrme model, we find that the spinning solutions are axially symmetric with energy proportional to the topological charge. There is also a maximal critical frequency above which solutions cease to exist. However, in the BPS baby Skyrme case we still have compacton solutions while here the size grows to infinity. Another difference is that here $\omega_{crit}$ is topological charge independent, while in the BPS baby model it goes as $\sqrt{n}$. Further, in the BPS model the size of the compactons in fact becomes smaller and smaller as we increase the frequency. Hence, in this aspect the Lifshitz model is more like the usual baby Skyrme model.   As a general conclusion, the different structure of (time-dependent) Lifshitz and Lorentz-symmetric field theories may lead to a qualitatively different behaviour of their time dependent solutions (for recent results on time-dependent solutions of Lorentz-violating nonlinear field theories we refer, e.g., to \cite{souza1}). 
%%%%%%%%%%%%%%%%%%%%%%%%%%%%%%%%%%%%%%%%%
\section{The gauged Lifshitz BPS baby Skyrme model}
%%%%%%%%%%%%%%%%%%%%%%%%%%%%%%%%%%%%%%%%%
The potential $V(\vec n \cdot \vec \phi)$ leaves an SO(2) $\simeq$ U(1) subgroup of the target space SO(3) symmetry intact, therefore a natural way to couple the Skyrme field to an electromagnetic gauge field is by gauging this residual U(1) symmetry. We shall require that the introduction of the gauge coupling via minimal substitution does not spoil the scaling properties of the theory, which implies that the gauge fields must scale under the anisotropic scaling like the corresponding partial derivatives. That is to say, $A_0$ should have scaling dimension $z$ (where, in our case, $z=d=2$), whereas the spatial components $A_j$ must have scaling dimensions equal to one. But this results in a Maxwell term $F_{\mu\nu}^2$ where different contributions have different scaling dimensions. Specifically, terms with time derivatives do not correspond to renormalizable interactions. In $d=2$ space dimensions, one way to avoid this problem is by introducing a Chern-Simons term instead of the Maxwell term. It may be checked easily that all contributions to the Chern-Simons term have scaling dimension $z+d=2d=4$ and, therefore, correspond to a scale-invariant, renormalizable interaction.  
%%%%%%%%%%%%%%%%%%%%%%%%%%%%%
\subsection{The model coupled to Chern-Simons electromagnetism}
%%%%%%%%%%%%%%%%%%%%%%%%%%%%%
The model we want to consider is the Lifshitz BPS baby Skyrme model of the last section, minimally coupled to Chern-Simons electromagentism. The corresponding lagrangian is
\begin{equation}
\mathcal{L}= \gamma (D_0 \vec{\phi})^2-\frac{\lambda^2}{2} Q^2 + \frac{\rho}{2}\epsilon^{\mu \nu \alpha} F_{\mu \nu}  A_\alpha - \mu^2 V(\vec{n}\cdot \vec{\phi} ) 
\end{equation}
where 
\begin{equation}
Q_{\mu\nu}= \vec{\phi} \cdot (D_\mu \vec{\phi} \times D_\nu \vec{\phi})=q_{\mu\nu} +A_\mu \partial_\nu (\vec{n}\cdot \vec{\phi}) - A_\nu \partial_\mu (\vec{n}\cdot \vec{\phi}) \equiv \epsilon_{\mu\nu\lambda}Q^\lambda,
\end{equation}
i.e.,
\begin{equation}
Q^\mu \equiv \frac{1}{2} \epsilon^{\mu\nu \lambda}Q_{\nu\lambda} = q^\mu + \epsilon^{\mu\nu\lambda} A_\nu \partial_\lambda (\vec n\cdot \vec \phi ),
\end{equation}
and $Q\equiv Q^0$.
Here, $\rho$ is a new coupling constant measuring the coupling of the Chern-Simons term, and $D_\mu$ is the usual covariant derivative
\begin{equation}
D_\mu \vec{\phi}=\partial_\mu \vec{\phi} + A_\mu \vec{n}\times \vec{\phi} .
\end{equation}
%%%%%%%%
\subsection{The BPS bound}
%%%%%%%%%%
Using the short-hand notation $Q\equiv Q^0$, $q\equiv q^0$, the lagrangian density may be re-expressed like
\begin{equation}
\mathcal{L}= \gamma \left( \dot{\vec \phi}{}^2 +2A_0 \dot{\vec \phi} \cdot (\vec n\times \vec \phi ) + A_0^2 \left( 1- (\vec n\cdot \vec \phi)^2 \right) \right) - \frac{\lambda^2}{2} Q^2 + \rho (A_0 B + \epsilon_{ij}E_i A_j) -\mu^2 V
\end{equation}
where we made the $A_0$ dependence explicit and introduced the electric and magnetic fields
\begin{equation}
E_i = \partial_0 A_i - \partial_i A_0 \; , \quad B=F_{12} = \partial_1 A_2 - \partial_2 A_1.
\end{equation}
We want to study the energy functional for static fields, but before this we have to solve the constraint equation (the Gauss law) of the model. The Gauss law is
\begin{equation}
\frac{\partial \mathcal{L}}{\partial A_0} - \partial_j \frac{\partial \mathcal{ L}}{\partial A_{0,j}} = 2\gamma \left( \dot{\vec \phi} \cdot (\vec n\times \vec \phi )  + A_0 \left( 1-(\vec n\cdot \vec \phi )^2 \right) \right) +2\rho B \equiv 0
\end{equation}
or, for static configurations
\begin{equation}
B=-\frac{\gamma}{\rho} A_0 \left( 1- (\vec n\cdot \vec \phi )\right) .
\end{equation}
The static lagrangian is
\begin{equation} \label{stat-lag}
\mathcal{L}= \gamma  A_0^2 \left( 1- (\vec n\cdot \vec \phi)^2 \right)  - \frac{\lambda^2}{2} Q^2 + \rho (A_0 B + \epsilon_{ij}E_i A_j) -\mu^2 V
\end{equation}
and, using 
\begin{equation} \label{tot-deriv-1}
\epsilon_{ij} E_i A_j = -\epsilon_{ij}(\partial_i A_0)A_j = -\partial_i (\epsilon_{ij} A_0 A_j) + A_0 B,
\end{equation}
skipping the total derivative, and eliminating $A_0$ via the Gauss law, we get the static energy density
\begin{equation} \label{stat-energy}
\mathcal{E} = - \mathcal{L} = \frac{\rho^2}{\gamma} \frac{B^2}{1-(\vec n\cdot \vec \phi )^2 } + \frac{\lambda^2}{2} Q^2 + \mu^2 V(\vec n\cdot \vec \phi)
\end{equation}
which is positive semi-definite. 
The total derivative $-\partial_i (\epsilon_{ij} A_0 A_j)$ which we omitted does not influence the Euler-Lagrange equations, but it might produce boundary contributions to the static energy, in which case the two expressions (\ref{stat-lag}) and (\ref{stat-energy}) for the static energies
$\int d^2 x\mathcal{E} = - \int d^2 x \mathcal{L}$ would give different results. We shall find, however, that BPS solutions always obey $\lim_{|\vec x|\to \infty} A_0 =0$, so the boundary term does not contribute and the two energy expressions provide the same energy.

Now we assume $\vec n = (0,0,1) \; \Rightarrow \; \vec n \cdot \vec \phi = \phi_3$ and use the new variable $h$ defined in (\ref{h-def}),
then the static energy functional is
 \begin{equation}
E  = \int d^2 x \left( \frac{\lambda^2}{2} Q^2 + \frac{ \rho^2}{4\gamma} \frac{B^2}{h(1-h) } + \mu^2 V(h) \right) .
\end{equation}
This is similar to the energy of the gauged BPS baby Skyrme model \cite{gaugedBPSbaby}, and also the BPS bound is similar. Indeed, we introduce a superpotential $W(h)$ and start with the non-negative expression
\begin{eqnarray}
0 &\le & \int d^2 x \left( \frac{\lambda^2}{2}(Q + W_h)^2 + \frac{\rho^2}{4\gamma} \frac{1}{h(1-h)} \left( B + \frac{4\gamma\lambda^2 }{\rho^2} h(1-h) W \right)^2 \right) \nonumber \\
&=& \int d^2 x \left[ \frac{\lambda^2 }{2} (Q^2 + W_h^2 ) + \frac{\rho^2}{4\gamma} \left( \frac{B^2}{h(1-h)} + \frac{16\gamma^2 \lambda^4 }{\rho^4 }h(1-h)W^2 \right) \right. \\
&& \left. + \lambda^2 \left( QW_h + 2BW \right) \right] .
\end{eqnarray}
Using $Q=q-2\epsilon_{ij} A_i \partial_j h$ the last line may be written as a topological term plus a total derivative,
\begin{equation} \label{tot-deriv-2}
QW_h +2BW = W_h q -2 \partial_j (\epsilon_{ij} A_i W) .
\end{equation}
We will find that $A_i$ does not vanish in the limit $\vec x \to \infty$, therefore the total derivative can be omitted only provided that $W$ obeys 
$\lim_{\vec x \to \infty}W=0$. But $W$ depends on $h$ only, and $h$ must take the vacuum value $h=0$ at infinity, therefore the condition which $W$ has to obey is 
\begin{equation} \label{W-bound-cond}
W(h=0)=0.
\end{equation}
We shall find that for the class of potentials $V(h)$ we consider in this paper, the corresponding superpotentials indeed always obey the above boundary condition. 

Assuming this, and comparing the non-negative expression with the energy, we find that the energy obeys the BPS inequality
\begin{equation}
E \ge \lambda^2 \vert \int d^2 x qW_h \vert
\end{equation}
provided that the superpotential $W$ obeys the modified superpotential equation
\begin{equation} \label{superpot-eq}
\frac{\lambda^2}{2} W_h^2 + \frac{4\lambda^4 \gamma}{\rho^2} h(1-h) W^2 = \mu^2 V
\end{equation}
together with the boundary condition (\ref{W-bound-cond}). Finally, the BPS bound is topological because $q$ is the pullback of the area two-form $d\Omega$ on the target space two-sphere S$^2$, and, therefore,
\begin{equation}
\int d^2 x qW_h = n \int d\Omega W_h =n\int_0^{2\pi} d\varphi \int_{-1}^1 d \phi_3 W_h = 4\pi n\int_0^1 dh W_h =4\pi n W(1) 
  \end{equation}
where $n$ is the winding number (degree) of the map defined by the Skyrme field configuration $\vec \phi$. 

The BPS bound is saturated for fields which obey the corresponding BPS equations
\begin{equation}
Q=  - W_h \; , \quad B= - \frac{4\gamma\lambda^2}{\rho^2 }h(1-h) W.
\end{equation}
It may be proved that the BPS equations imply the static Euler-Lagrange equations. For general static configurations, the proof is rather lengthy and is similar to the proof in \cite{gaugedBPSbaby}. Here we shall give the explicit proof for axially symmetric configurations in the next section.

We remark that, in principle, there exists a further term with Lifshitz scaling dimension equal to $z+d=2d=4$ which we might include, namely the gauge potential coupled (non-minimally) to the topological current. The corresponding contribution to the action will be gauge invariant only provided that the current is both conserved and gauge invariant, so neither $q^\mu$ (which is not gauge invariant) nor $Q^\mu$ (which is not conserved) may be used. There exists, however, a gauge invariant and conserved topological current, namely $\tilde Q^\mu =Q^\mu +(1/2) \epsilon^{\mu \nu \rho} F_{\nu \rho} (1-\vec{n}\cdot \vec{\phi})$ \cite{schroers1}, therefore the term $\tilde \rho \int d^2 x dt \tilde Q^\mu A_\mu$ is gauge invariant. The resulting theory with this term included still has a BPS bound and BPS equations in terms of a superpotential, and the structure of the superpotential equation is similar (first order and quadratic). Concretely, the superpotential equation has the form
$ AW'^2 +2BWW' + CW^2 = V$, where $A,B,C$ are (quite complicated) rational functions of $h$ and define a non-negative quadratic form, i.e., $AC-B^2 \ge 0$. BPS soliton solutions could then be found, e.g., by numerically integrating the resulting three first order equations (the two BPS equations and the superpotential equation). For reasons of simplicity, however, we will not further consider this general case here and set $\tilde \rho =0$ in the rest of the paper.  
%%%%%%%%%%%%%%%%%%%%%%%%%%%%%
\subsection{The axially symmetric ansatz}
%%%%%%%%%%%%%%%%%%%%%%%%%%%%%
Next we assume $\vec n = (0,0,1)$ and the standard static ansatz
\begin{equation} \label{rad-ans}
\vec{\phi} (r,\phi)  = \left( 
\begin{array}{c}
\sin f(r) \cos n\phi \\
\sin f(r) \sin n\phi \\
\cos f(r)
\end{array}
\right), \;\;\;\; A_0=nb(r), \;\;\; A_r=0, \;\;\; A_\phi=na(r) .
\end{equation}
Then the magnetic field is 
\begin{equation}
B=\frac{1}{2} \epsilon^{0ij} F_{ij}= \epsilon^{0ij} \partial_i A_j= \frac{na'(r)}{r} .
\end{equation}
The resulting Lagrangian is
\begin{equation}
L=
2\pi n^2 \int rdr \left(\gamma b^2 \sin^2 f -\frac{\lambda^2}{4} 2 \sin^2f \frac{f_r^2}{r^2} (1+a)^2 +  \frac{\rho}{r}( a_rb -  b_r  a) -\frac{\mu^2}{n^2} V(f) \right) .
\end{equation}
We introduce the variable $x=r^2 /2$ and the new target space variable $h$ defined in (\ref{h-def}), where
\begin{equation}
h = \frac{1}{2}(1-\cos f) \quad \Rightarrow \quad h_x = \frac{1}{2} \sin f f_x ,
\end{equation}
then
\begin{equation}
L=2\pi n^2 \int dx \left( 4\gamma b^2h(1-h)-2 \lambda^2 h_x^2 (1+a)^2 + \rho ( a_x b-  b_x  a) -\frac{\mu^2}{n^2} V(h) \right) .
\end{equation}
The static field equations read
\begin{equation} \label{axi-h-eq}
\partial_x \left( h_x (1+a)^2 \right)  - \frac{\mu^2}{4\lambda^2 n^2} V_h+\frac{\gamma}{\lambda^2} b^2(1-2h)=0
\end{equation}
\begin{equation} \label{axi-b-eq}
b_x=-\frac{2\lambda^2}{\rho}h_x^2(1+a) ,
\end{equation}
and the Gauss law is
\begin{equation} \label{axi-gauss}
a_x= -\frac{4 \gamma}{\rho} b h(1-h) .
\end{equation}
Using the Gauss law to eliminate $b$ we get the positive definite Hamiltonian
\begin{equation} \label{spher-hamilton}
H=-L=2\pi n^2 \int dx \left( 2 \lambda^2 h_x^2 (1+a)^2 +\frac{\rho^2}{4\gamma} \frac{a_x^2}{h(1-h)}+\frac{\mu^2}{n^2} V(h) \right) .
\end{equation}
In a next step, let us prove that, for the axially symmetric ansatz, the BPS equations together with the superpotential equation and the Gauss law imply the static second order equations. The BPS equations for the axially symmetric ansatz read
\begin{eqnarray} \label{axi-1-bps-eq}
2nh_x (1+a) &=&-W_h \\  \label{axi-2-bps-eq}
na_x &=& -\frac{4\gamma\lambda^2}{\rho^2} h(1-h) W .
\end{eqnarray}
We remark that the superpotential equation (\ref{superpot-eq}) is invariant under the sign change $W\to -W$. We chose the sign such that $W$ and $W_h$ will be positive in general. This follows from the fact that $h_x$ is negative, in general, because $h(0)=1$, $h(\infty)=0$.

Using the second BPS equation (\ref{axi-2-bps-eq}) and the axially symmetric Gauss law (\ref{axi-gauss}) to eliminate $a_x$ we easily find
\begin{equation} \label{b-W-eq}
b=\frac{\lambda^2}{n\rho} W \quad \Rightarrow \quad b_x = \frac{\lambda^2}{n\rho} W_h h_x = -2\frac{\lambda^2}{\rho} h_x^2 (1+a),
\end{equation} 
i.e., Eq. (\ref{axi-b-eq}), where we used the first BPS equation (\ref{axi-1-bps-eq}) in the last step. Finally, the field equation (\ref{axi-h-eq}) may be transformed into the $h$ derivative of the superpotential equation by the following series of steps,
\begin{eqnarray}
\partial_x \left( h_x (1+a)^2 \right) + \frac{\gamma}{\lambda^2} b^2 (1-2h) &=& \frac{\mu^2 }{4n^2 \lambda^2}V_h \nonumber \\
-\frac{1}{2n} \partial_x \left( W_h (1+a)\right) + \frac{\gamma \lambda^2}{n^2 \rho^2} W^2 (1-2h) &=& \frac{\mu^2 }{4n^2 \lambda^2}V_h \nonumber \\
-\frac{1}{2n} \left( W_{hh} h_x (1+a) + W_h a_x \right) + \frac{\gamma \lambda^2}{n^2 \rho^2} W^2 (1-2h) &=& \frac{\mu^2 }{4n^2 \lambda^2}V_h
\nonumber \\
\frac{1}{2n^2}\left( \frac{1}{2} W_{hh} W_h + \frac{4\gamma\lambda^2}{\rho^2} h(1-h) W W_h \right) 
+ \frac{\gamma \lambda^2}{n^2 \rho^2} W^2 (1-2h) &=& \frac{\mu^2 }{4n^2 \lambda^2}V_h
\nonumber \\
\partial_h \left( \frac{\lambda^2}{2} W_h^2 + \frac{4\gamma \lambda^4}{\rho^2} h(1-h) W^2 \right) &=& \mu^2 V_h
\end{eqnarray}
where the last line is just the $h$ derivative of the superpotential equation, and we used the BPS equations and the Gauss law in several instances. 
Reading the above derivation from below to above, it follows that also the first field equation (\ref{axi-h-eq}) is a consequence of the BPS equations and the superpotential equation.
%%%%%%%%%%%%%%%%
\subsection{The condition $W(h=0)=0$}
%%%%%%%%%%%%%%%%%
Now, let us demonstrate that for the class of potentials we want to consider the corresponding superpotential which gives rise to the BPS solutions must always obey the boundary condition $W(h=0)=0$. The class of potentials we permit have their vacuum at $h=0$, i.e., $V(h=0)=0$. We do not exclude the possibility of more vacua, but it is always the vacuum at $h=0$ which is approached in the limit of large $|\vec x|$, i.e. $\lim_{|\vec x|\to \infty }h(\vec x)=0$. Further, we assume that the vacuum at $h=0$ is approached like
\begin{equation}
V(h) \sim h^\alpha \; , \quad \alpha \ge 1.
\end{equation}
For $\alpha >1$, the condition $W(0)=0$ is a direct consequence of the superpotential equation (\ref{superpot-eq}). Indeed, the superpotential equation near $h=0$ implies that $W_h \sim h^{(\alpha /2)}$, which a priori seems to allow for the superpotential $W \sim h^{(\alpha /2) +1} + W_0$ where $W_0$ is a constant. But let us study the $h$ derivative of the superpotential equation near $h=0$, 
\begin{equation} \label{superpot-der-eq}
\lambda^2 W_h W_{hh} + \frac{4\lambda^4 \gamma}{\rho^2} ((1-2h)W^2 + 2h(1-h) WW_h ) = \mu^2 \alpha h^{\alpha -1}.
\end{equation}
For $\alpha >1$, both the r.h.s. of the above equation and the product $W_hW_{hh}$ are zero at $h=0$, which implies $W(h=0)=0$, which is what we wanted to prove. 

The case $\alpha =1$, on the other hand, is more involved. It is, at the same time, important, because it includes well-known potentials like the "old" and "new" baby Skyrme potentials. In the above equation (\ref{superpot-der-eq}), both $W_hW_{hh}$ and $h^{\alpha -1}$ give finite, nonzero contributions at $h=0$ for $\alpha =1$, and the superpotential equation allows, in fact, a one-parameter family of solutions parametrized by the boundary value $W_0 \equiv W(0)$. We shall demonstrate, however, that only the case $W_0 =0$ leads to a genuine soliton solution with finite energy, whereas other choices lead to formal solutions with singularities or infinite energy. For simplicity, we will restrict to the case of spherically symmetric field configurations $h(x)$, $a(x)$, i.e., to the variational problem defined by the energy functional (\ref{spher-hamilton}). It turns out that the discussion is simpler directly for the second order system resulting from this energy, where the relation to the first order (BPS) system will be obvious. The field equations resulting from (\ref{spher-hamilton}) are (equivalently they may be derived from Eqs. (\ref{axi-h-eq}), (\ref{axi-b-eq}) and (\ref{axi-gauss}) by eliminating $b$ with the help of the Gauss law)
\begin{equation} \label{axi-h-eq2}
4 \lambda^2 \partial_x(h_x (1+a)^2) + \frac{1-2h}{h^2(1-h)^2} \frac{\rho^2}{4 \gamma} a_x^2 = \frac{ \mu^2}{n^2} V_h,
\end{equation}
\begin{equation} \label{axi-a-eq2}
4 \lambda^2 h_x^2(1+a) = \frac{\rho^2}{2 \gamma} \partial_x \left( \frac{a_x}{h(1-h)} \right).
\end{equation}
As a next step, we need the information that for $V\sim h$ the solitons are compactons which approach their vacuum values at a finite distance $x_0$. Concretely, the approach is quadratic, i.e., like $h\sim h_2 (x_0 -x)^2$. As a consequence, the boundary conditions for the variational problem are
\begin{equation}
h(0) = 1, \quad  a(0) = 0, \quad
h(x_0) = 0, \quad  h'(x_0) = 0, \quad  a'(x_0) = 0.
\end{equation}
On the other hand, we have a total number of five free constants (four integration constants and the compacton radius $x_0$), so free constants and boundary conditions seem to match exactly, leading to at most one solution. This conclusion is, however, not correct. The problem is that the condition $a'(x_0)=0$ is not an independent condition but, instead, a consequence of the field equations and the remaining boundary conditions. So apparently we found a one-parameter family of solutions (five free parameters and four independent boundary conditions). The problem with this family of apparent solutions is that different solutions will, in  general, have different energies. Indeed, for a fixed value of the compacton radius $x_0$ the number of free parameters and boundary conditions match, so we expect at most one solution. Different values of $x_0$, however, correspond to different variational problems (the same equations and boundary conditions, but on different intervals), so there is no reason why the energies should be the same. This would imply that the original variational problem is ill-defined. The resolution of the puzzle is that we have to impose one more condition for a genuine compacton. A genuine compacton is not a solution on the finite interval $[0,x_0]$. Instead, it is a solution on the full half-line which allows to connect a nontrivial solution for $x\in [0,x_0]$ with the vacuum solution for $x\ge x_0$. The most direct way to derive this additional condition is by inserting the power series expansion
\begin{equation}
h = \sum_{k=2}^\infty h_k (x_0 -x)^k 
\end{equation}
\begin{equation}
a = \sum_{k=0}^\infty a_k (x_0 - x)^k
\end{equation}
into the field equations (\ref{axi-h-eq2}), (\ref{axi-a-eq2}). As we are expanding near the vacuum $h=0$, we replace the potential $V$ by $V\to h$ and $V_h$ by $V_h \to 1$. We find that $a_0$ remains undetermined, whereas $a_1 =0$ and $a_2 =0$. The next condition is
\begin{equation}
\frac{9 \rho}{4 \gamma} a_3^2 + h_2^2 \left( -\frac{ \mu^2}{n^2} + 8\lambda^2(1+a_0)^2 h_2 \right) = 0.
\end{equation} 
This condition should be interpreted as a determining equation for $h_2$ for given values of $a_0$ and $a_3$. It is a cubic equation in $h_2$, so it will, in general, lead to three solutions (roots) for $h_2$. It will, therefore, allow to join different solutions at $x_0$, corresponding to the choice of different roots $h_2$. The vacuum solution $h_2=0$, however, may be found among the roots only for $a_3 =0$. For a compacton solution we must, therefore, require the additional condition $a_3 =0$. For this choice we may join the nontrivial solution
\begin{equation} \label{h-2}
h_2 = \frac{\mu^2}{8n^2 \lambda^2 (1+a_0)^2} 
\end{equation}
for $x\le x_0$ with the vacuum solution $h_2 =0$ for $x>x_0$. In addition, we have now two free constants ($a_0$ and $x_0$) to satisfy the two remaining boundary conditions $h(0)=1$ and $a(0)=0$ at the center $x=0$, so we expect at most one solution, and, therefore, a well-defined variational problem.

Finally, it easily follows from the BPS equation Eq. (\ref{axi-2-bps-eq}) that $a_3=0$ implies $W(h=0)=0$ which, as a consequence, continues to hold also for potentials which are linear in $h$ near the vacuum $h=0$. It further follows from Eq.  (\ref{b-W-eq}) that $W(h=0)=0$ implies $b(x_0)=0$ (or $\lim_{x\to \infty} b(x)=0$ for non-compact solitons) and, therefore, $\lim_{|\vec x| \to \infty} A_0 (\vec x)=0$. As a result, the total derivative terms in Eqs. (\ref{tot-deriv-1}) and (\ref{tot-deriv-2}) do not contribute to the energy, as announced.

%%%%%%%%%%%%%%%%%%%%%%%%%%%%%
\subsection{The magnetic flux}
%%%%%%%%%%%%%%%%%%%%%%%%%%%%%
It is possible to find an exact expression (i.e., in terms of the superpotential $W$) for the magnetic flux
\begin{equation}
\Phi = \int r dr d\phi B = 2\pi n \int dx a_x = 2\pi n a(x_0) \equiv 2\pi n a_\infty
\end{equation}
where $x_0$ is the position of the boundary of the soliton which can be finite in the case of compactons as well as infinite for the usual exponentially or power-like localized solutions.  From the BPS equations for the axially symmetric ansatz we get
\begin{equation}
\frac{a_x}{1+a} = \frac{8\gamma \lambda^2}{\rho^2} \frac{h(1-h)W}{W_h} h_x,
\end{equation}
which can be integrated to
\begin{equation}
\partial_x \ln (1+a) = \frac{8\gamma \lambda^2}{\rho^2} \partial_x F(h)
\end{equation}
where
\begin{equation}
F_h \equiv \frac{h(1-h)W}{W_h} \;\;\; \Rightarrow \;\;\; F(h)= \int_0^h dh' \frac{h'(1-h')W(h')}{W_{h'}(h')} 
\end{equation}
It gives
\begin{equation} \label{1+a-eq}
 \ln C(1+a) =  \frac{8\gamma \lambda^2}{\rho^2} F(h(x)) \quad \Rightarrow \quad 1+a(x)= C^{-1}e^{\frac{8\gamma \lambda^2}{\rho^2}F(h(x))}
\end{equation}
where the integration constant $C$ can be eliminated using the boundary conditions $h(x=0)=1$ and $a(x=0)=0$. Then, 
\begin{equation}
\ln C =   \frac{8\gamma \lambda^2}{\rho^2} F(1) \;\;\; \Rightarrow \;\;\; C = e^{ \frac{8\gamma \lambda^2}{\rho^2} F(1)}
\end{equation}
Hence finally 
\begin{equation} \label{a-infty}
a_\infty = -1 + e^{ \frac{-8\gamma \lambda^2}{\rho^2} F(1)}
\end{equation}
where $h(x_0)=0$ and $F(h=0)=0$ have been used. As we see, the final expression is quite similar to the case of the gauged BPS baby case with the Maxwell gauge field. However, the specific form of the function $F(h)$ is different which could potentially lead to different properties as, e.g., quantization of the magnetic flux.
\\
The quantization of the flux would appear if the function $F$ tends to infinity at $h=1$ which obviously requires that the integrand has a singularity $(1-h)^{-1}$ or stronger. The simplest realization of this case is given by the superpotential $W$ having the following properties: \\
(i) $W(0)=0$ and no other zeros for $W$
\\
(ii) $W_h \sim (1-h)^a$, where $a \geq 2$ as $h \rightarrow 1$.
\\
As an example one may consider the superpotential $W$ in the following form
\begin{equation}
W=\frac{h^3}{3}-\frac{h^4}{2} +\frac{h^5}{5}=h^3\left( \frac{1}{3} -\frac{h}{2}  +\frac{h^2}{5} \right)
\end{equation} 
which has only one zero for $h=0$ on the segment $h \in [0,1]$. Then
\begin{equation}
W_h=h^2(1-h)^2 .
\end{equation} 
From the superpotential equation we may easily find the potential $V$ corresponding to this choice. It is a well behaving potential with two vacua located at the ends of the segment. Now
\begin{equation}
F_h= \frac{h^2}{1-h} \left( \frac{1}{3} -\frac{h}{2}  +\frac{h^2}{5} \right)
\end{equation} 
which is singular at $h=1$. It means that $F(1)=\infty$ and 
\begin{equation}
a_\infty = -1 .
\end{equation} 
So, at least for this potential $V$ we get {\it quantization} of the magnetic flux for all values of the coupling constants. However, one must check whether this superpotential leads to solitonic solutions with the prescribed boundary conditions. We will see that this is not the case.
\\
It can be easily shown from the first BPS equation (\ref{axi-1-bps-eq}). In the vicinity of the second vacuum $h=1$ (at $x=0$) the gauge field vanishes and we arrive at the simple equation 
\begin{equation}
2nh_x \sim (1-h)^{a-1}
\end{equation}
where we used the superpotential given above. The pertinent solution at the origin reads
$$ 1-h \sim \frac{1}{x^{1/(a-2)}}.$$
Hence, $h=1$ can be reached only for $x \rightarrow \infty$, which is incompatible with the boundary condition at the origin. In other words, the only solution with our boundary conditions is the trivial (vacuum) solution. Therefore, the flux cannot be quantized. We remark that in this derivation we only needed the behaviour of $W_h$ near $h=1$, which, in turn, was necessary for the condition $F(1)=\infty \Rightarrow a_\infty =-1$, so the impossibility of flux quantization seems to be a general result.
\\
As usual for the Chern-Simon type models, the magnetic flux is related to the electric charge. To see it, let us write the field equation for the gauge field as
\begin{equation}
2\rho \epsilon^{\mu \nu \sigma} \partial_\nu A_\sigma = -j^\mu
\end{equation}
where the current $j^\mu$ contains only the gauge field and not its derivatives. Obviously the current is conserved $\partial_\mu j^\mu=0$. Knowing that
\begin{equation}
B=-\frac{1}{2\rho} j^0, \;\;\; E_i=\frac{1}{2\rho} \epsilon_{ik}j^k
\end{equation}
we get
\begin{equation}
\Phi = \int d^2 x B=-\frac{1}{2\rho} \int d^2 x j^0=-\frac{Q}{2\rho}
\end{equation}
where $Q$ is the electric charge. 
%%%%%%%%%%%%%%%%%%%%%%%%%%%%%
\subsection{Gradient flow}
%%%%%%%%%%%%%%%%%%%%%%%%%%%%%
Here we want to demonstrate that the BPS equations for the axially symmetric ansatz may be expressed as a gradient flow equation w.r.t. a certain target space metric, exactly like in the cases where the superpotential and the corresponding BPS equations are derived via "fake supergravity" for self-gravitating domain walls, inflaton scalar fields, or extremal black holes with some (hyper-)spherical symmetry \cite{susy1}-\cite{f susy}.
Let us start with the BPS equations in the axially symmetric ansatz
\begin{eqnarray}
2nh_x (1+a) &=& -W_h \\  %\label{axi-2-bps-eq}
na_x &=& -\frac{4\gamma\lambda^2}{\rho^2} h(1-h) W 
\end{eqnarray}
Now, we introduce a new target space variable $a+1=e^{-U}$ and a new superpotential $\mathcal{W} = e^U W(h)$. Then we arrive at
 \begin{eqnarray}
h_x &=& -\frac{1}{2n} \mathcal{W}_h \\ % \label{axi-2-bps-eq}
U_x &=& \frac{4\gamma\lambda^2}{n\rho^2} h(1-h) \mathcal{W}_U 
\end{eqnarray}
One can write such BPS equations as first order gradient flow equations
\begin{equation}
\dot{q}^a-G^{ab} \frac{\partial \mathcal{W}}{\partial q^b}=0
\end{equation} 
where $q^a=(U,h)$ are the target space coordinates. The "effective metric" on the target space is
\begin{equation}
G_{ab}=\left(
\begin{array}{cc}
\frac{n\rho^2}{4\gamma\lambda^2} \frac{1}{h(1-h)}  & 0\\
 0&  -2n
\end{array}
\right) .
\end{equation}
As in the case of the BPS baby Skyrme model coupled to the Maxwell gauge field, we find a Minskowskian signature of the gradient flow target space. This resembles the case of extremal black holes but is different from the cases of domain walls and inflaton fields, which lead to an Euclidean signature. However, our metric is not constant. It becomes singular at the north and south poles of the $S^2 \ni \vec{\phi}$. 
We remark that a detailed discussion of gradient flows and the related supertpotentials in Lifshitz field theories can be found in \cite{bakas2} where, however, the flow is parametrized by euclidean time and not by a space variable (like $x=r^2/2$, in our case).
%%%%%%%%%%%%%%%%%%%%%%%%%%%%%
\subsection{Exactly solvable superpotential equation}
%%%%%%%%%%%%%%%%%%%%%%%%%%%%%
For the new baby potential $V=h(1-h)$ we can solve the superpotential equation completely. We have 
\begin{equation}
\frac{\lambda^2}{2} W_h^2 + \frac{4\lambda^4 \gamma}{\rho^2} h(1-h) W^2 = \mu^2 h(1-h)
\end{equation}
\begin{equation}
\frac{\lambda^2}{2} W_h^2 =  h(1-h) \mu^2 \left(1- \frac{4\lambda^4 \gamma}{\mu^2 \rho^2} W^2\right)
\end{equation}
\begin{equation}
\frac{\lambda}{\sqrt{2} \mu} \frac{dW}{\sqrt{1- \frac{4\lambda^4 \gamma}{\mu^2 \rho^2} W^2}} = \pm  \sqrt{h(1-h)} dh 
\end{equation}
Hence, 
\begin{equation}
\frac{\rho}{2\sqrt{2} \lambda \sqrt{\gamma}} \arcsin \left( \frac{2\lambda^2\sqrt{\gamma}}{\mu \rho} W\right)=  \pm \frac{1}{8} \left[ 2(1-2h)\sqrt{h(1-h)} +\arcsin (1-2h) +C \right]
\end{equation}
or
\begin{equation}
W=  \pm \frac{\mu \rho}{2\lambda^2\sqrt{\gamma}} \sin \left( \frac{\lambda \sqrt{\gamma}}{2\sqrt{2} \rho} \left[ 2(1-2h)\sqrt{h(1-h)} +\arcsin (1-2h) +C \right] \right)
\end{equation}
where $C$ is an integration constant fixing different $W(0)$ values. We already know that $W(0)=0$ is the only acceptable value, which implies $C=-(\pi /2)$.

From the explicit expression above it is obvious that a superpotential with the right boundary condition $W(0)=0$ exists for arbitrary values of the coupling constants. This does not imply, however, that BPS soliton solutions will exist for all values of the couplings. Indeed, for sufficiently large values of $\frac{\lambda\sqrt{\gamma}}{2\sqrt{2}\rho}$, the derivative $W_h$ will have a zero in the interval $h\in (0,1)$, and in this case a BPS soliton does not exist. The argument goes like follows. Let us assume that $W_h$ has exactly one zero in the open interval $h\in (0,1)$. Then $W_h$ is either positive below the zero and negative above, or vice versa. In any case, it has different signs near $h=0$ and near $h=1$. But via the BPS equation $2nh_x(1+a)=-W_h$ this implies that $h_x$ has different signs near $x=0$ and near $x=\infty$, which is incompatible with the conditions $h(0)=1$, $h(\infty)=0$ and $h\in [0,1]$. In this argument we used that $1+a(x)\ge 0 \; \forall \; x$, as follows from Eq. (\ref{1+a-eq}).  
This argument does not exclude the possiblity of BPS solitons for superpotentials such that $W_h$ has an even number of zeros in the interval $(0,1)$. 
But for the exact superpotential of this section, this last possibility can be excluded, too. 
In order to see it, we may consider the expression for the total flux which is completely fixed by the value of the previously introduced function $F(h)$ at $h=1$ (see Section IV.D). In the case of the exact superpotential we get
\begin{eqnarray}
F(1)= \frac{1}{8 \kappa} \int_0^1 dh \sqrt{h(1-h)} \tan \left( \kappa \left[ 2(1-2h)\sqrt{h(1-h)} +\arcsin (1-2h) -\frac{\pi}{2} \right] \right) \\
=\frac{1}{8 \kappa} \int_0^1 dh \sqrt{h(1-h)} \tan \left( \kappa\; p(h) \right) 
\end{eqnarray}
where $\kappa=\lambda \sqrt{\gamma}/ (2\sqrt{2} \rho)$ and $p(h)= 2(1-2h)\sqrt{h(1-h)} +\arcsin (1-2h) -\frac{\pi}{2}$. Surprisingly the integral is analytical as the derivative of the argument of tangent function is proportional to the overall factor i.e., $\frac{dp(h)}{dh}=-8\sqrt{h(1-h)}$. Finally,  
\begin{equation}
F(1)= -\frac{1}{64 \kappa^2} \ln \cos \kappa \pi 
\end{equation}
which is finite only for 
\begin{equation}
\cos \kappa \pi > 0 \;\;\; \Rightarrow \;\;\;  \kappa  < \frac{1}{2}
\end{equation}
Hence, for $\kappa >(1/2)$ the flux is not well defined on the whole segment $h \in [0,1]$ which, in consequence, excludes such solutions.  
\\
For $\kappa =1/2$, we get that $F(1)=\infty$ which leads to flux quantization and is not compatible with the boundary conditions for a soliton, as was discussed in Section IV.D. Solitons may, therefore, only exist for $\kappa < (1/2)$. Let us notice that for solutions corresponding to the choice of the model parameters approaching the critical value $\kappa =1/2$ we asymptotically (with arbitrary accuracy) tend to the quantized magnetic flux. The approach is rather weak (quadratical) as
\begin{equation}
a_\infty=-1+ \cos \kappa \pi
\end{equation}
which should be contrasted with typical exponential (or even more rapid) approach for the large gauge coupling limit in the gauged (Maxwell) BPS baby Skyrme model. Interestingly enough, the asymptotic quantization of the flux occurs for finite values of the parameters of the model which again differs from the BPS baby Skyrme model with the Maxwell field, where such a limit is realized only for infinite value of the gauge coupling constant. 
\\
We remark that for generic potentials, the occurrence of zeros of the superpotential derivative, $W_h=0$, at some point $h_0 \in (0,1)$ implies that $W_{hh}$ becomes singular at this point, and the superpotential equation cannot be integrated beyond this point. In these cases, BPS solitons obviously do not exist. Only for some specific potentials (like the one studied in this section), the superpotential exists in the whole interval even if the interval contains points where $W_h=0$, and the possible existence of BPS solitons requires some further study.  

%%%%%%%%%%%%%%
\subsection{Numerical results}
In a next step, we want to numerically integrate the field equations for the axially symmetric ansatz, where we integrate directly the system of three equations (\ref{axi-h-eq}), (\ref{axi-b-eq}) and (\ref{axi-gauss}) for $h$, $a$ and $b$. Concretely, we integrated the system for the following three potentials. 
\subsubsection{The old potential $V=2h$} 
\begin{figure}[h]
 \begin{center}
  \subfloat[Function $h$ with its derivative.]{\includegraphics[width=0.45\textwidth]{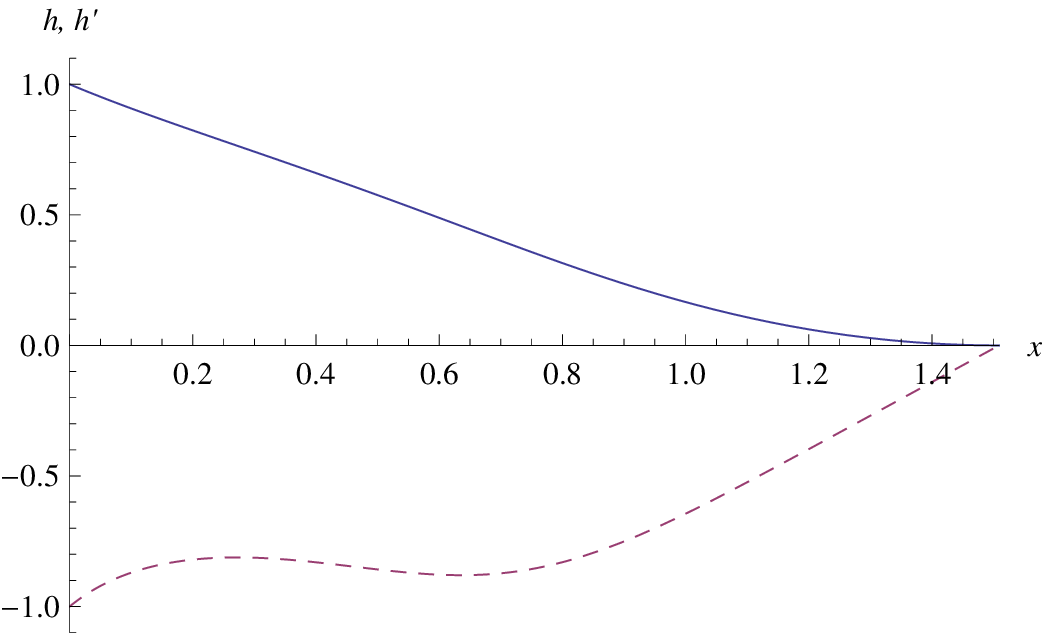}} \quad
  \subfloat[Function $a$ and its derivative.]{\includegraphics[width=0.45\textwidth]{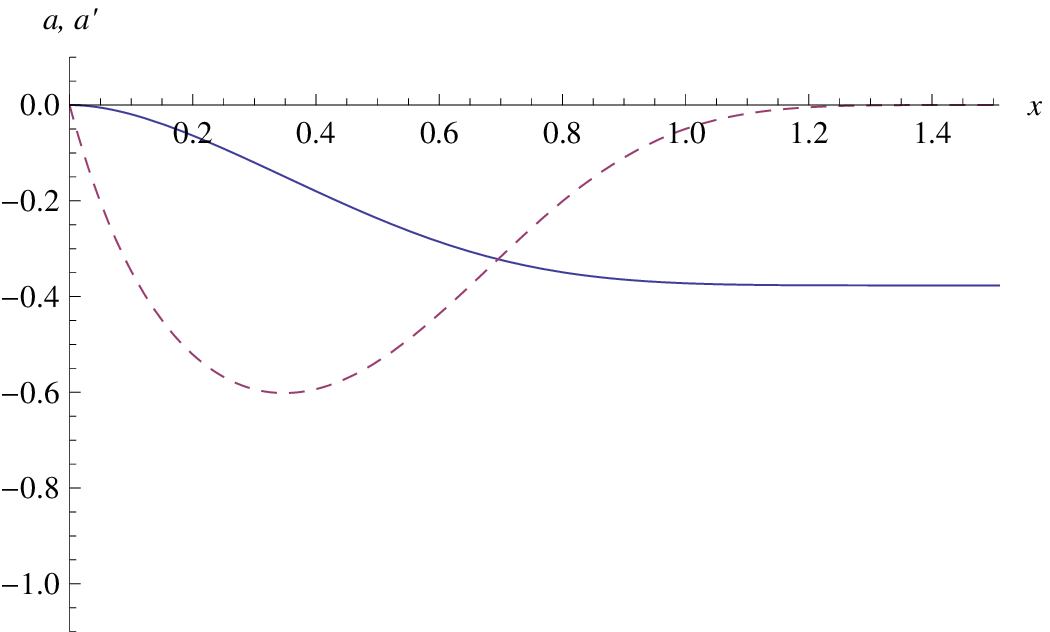}} \\
  \subfloat[Function $b$ and its derivative.]{\includegraphics[width=0.45\textwidth]{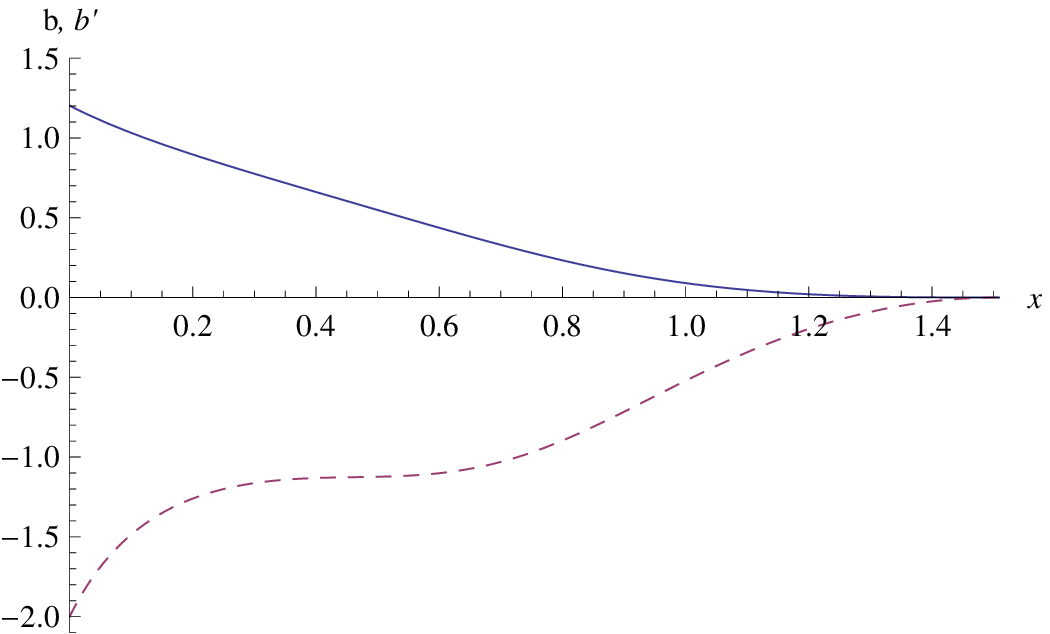}} \quad
  \subfloat[Energy density.]{\includegraphics[width=0.45\textwidth]{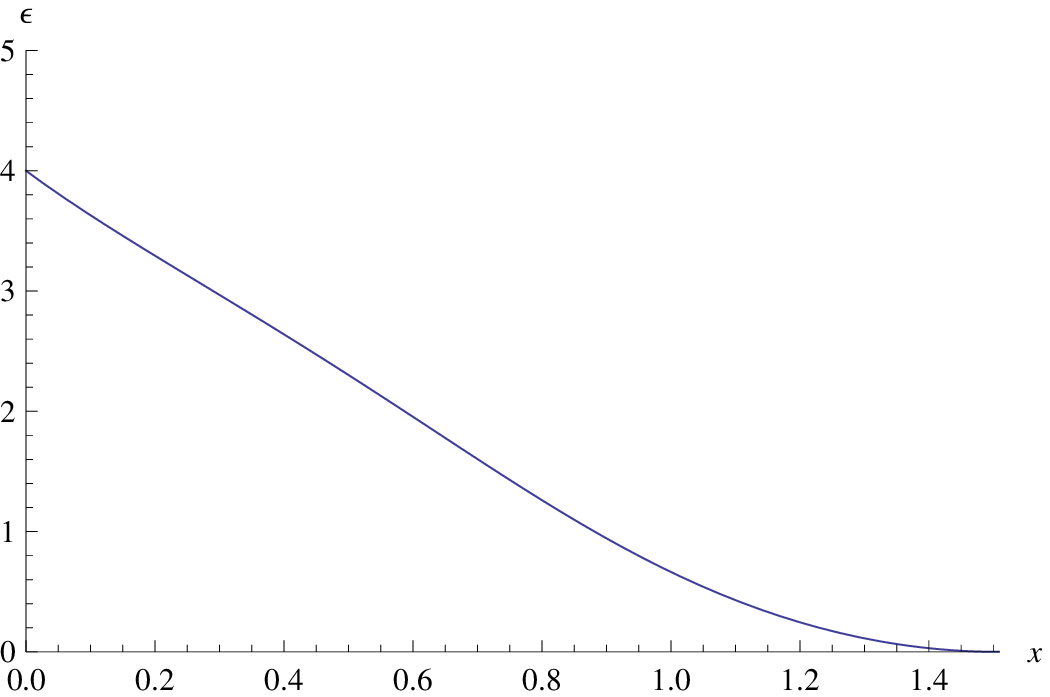}} \\
  \caption{Static solution for the old baby potential, with the coupling constant values $\mu = \gamma =\lambda = \rho = n = 1$.}
  \label{Old}
 \end{center}
\end{figure}

Solutions are compactons, and we perform a shooting from the compacton boundary. For the correct boundary conditions at the boundary $x_0$, the expansions read
\begin{equation}
h(x) \sim \frac{\mu^2}{4n^2 \lambda^2 (1+a_0)^2} (x_0 -x)^2 + O(x_0 -x)^4,
\end{equation}
\begin{equation}
a(x) \sim a_0  + O(x_0 -x)^5,
\end{equation}
\begin{equation}
b(x) \sim \frac{\mu^4}{96 \rho n^4 \lambda^2 (1+a_0)^3}  (x_0 -x)^3 + O(x_0 -x)^5,
\end{equation}
(the coefficient $h_2$ here is a factor of 2 bigger than in Eq. (\ref{h-2}), because here $V=2h$, whereas we assumed $V\sim h$ in Section III.D). The free constants are $a_0$ and $x_0$, and they must be determined such that the two conditions $h(0)=1$ and $a(0)=0$ are satisfied. The result of the numerical integration for a specific choice of the coupling constants is shown in figure \ref{Old}. The values of the free constants and the energy for this solution are
\begin{equation}
 a_0 = -0.376724,  \qquad x_0 = 1.5089, \qquad E_0 = 15.1067.
\end{equation}
Further, the energy coincides exactly with the BPS energy, and $a_0$ is exactly equal to the value given by Eq. (\ref{a-infty}), which demonstrates that this solution is, in fact, a BPS solution. \\
\subsubsection{The new potential $V=h(1-h)$} 
\begin{figure}[h]
 \begin{center}
  \subfloat[Function $h$ with its derivative.]{\includegraphics[width=0.45\textwidth]{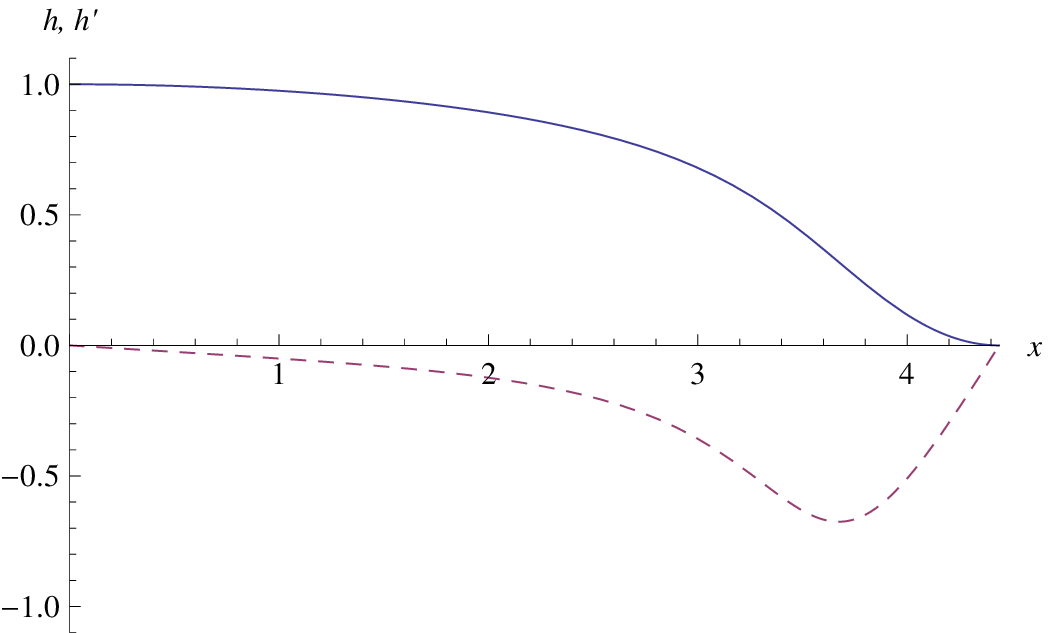}} \quad
  \subfloat[Function $a$ and its derivative.]{\includegraphics[width=0.45\textwidth]{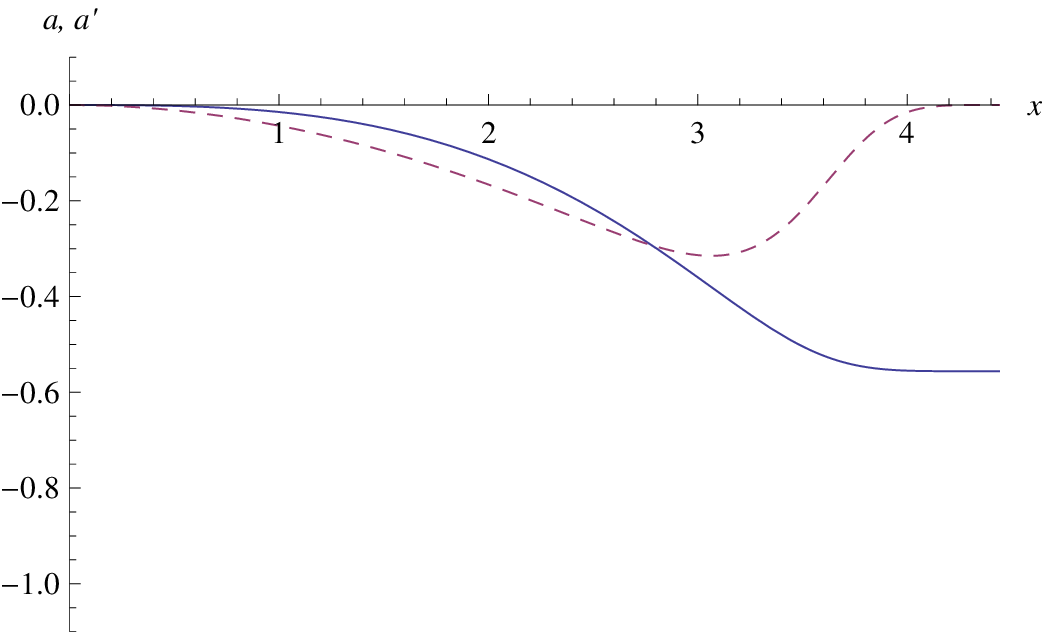}} \\
  \subfloat[Function $b$ and its derivative.]{\includegraphics[width=0.45\textwidth]{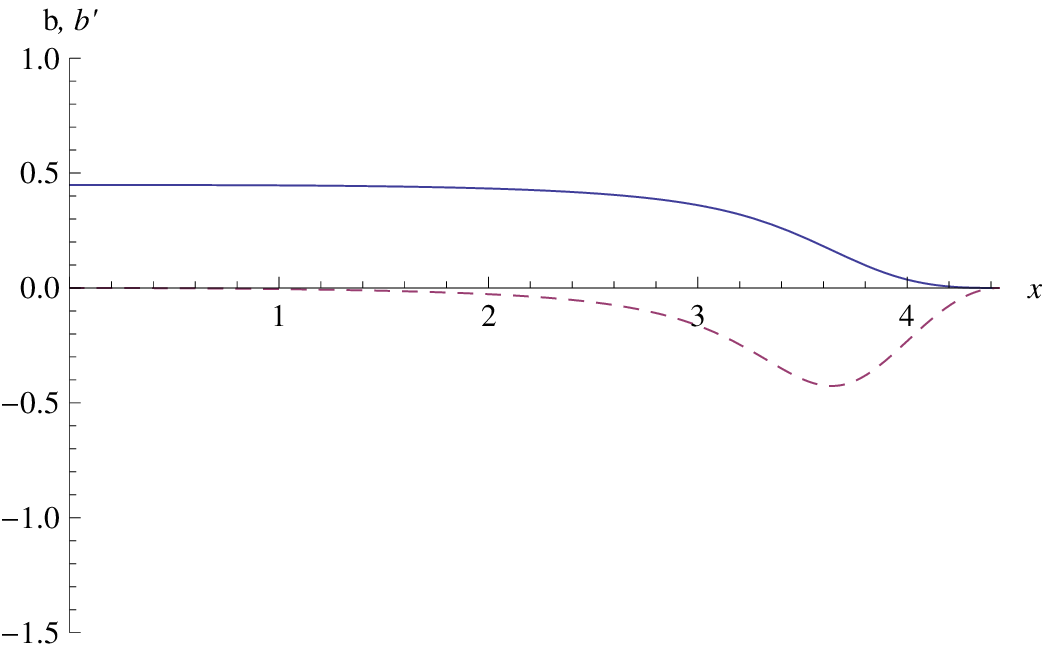}} \quad
  \subfloat[Energy density.]{\includegraphics[width=0.45\textwidth]{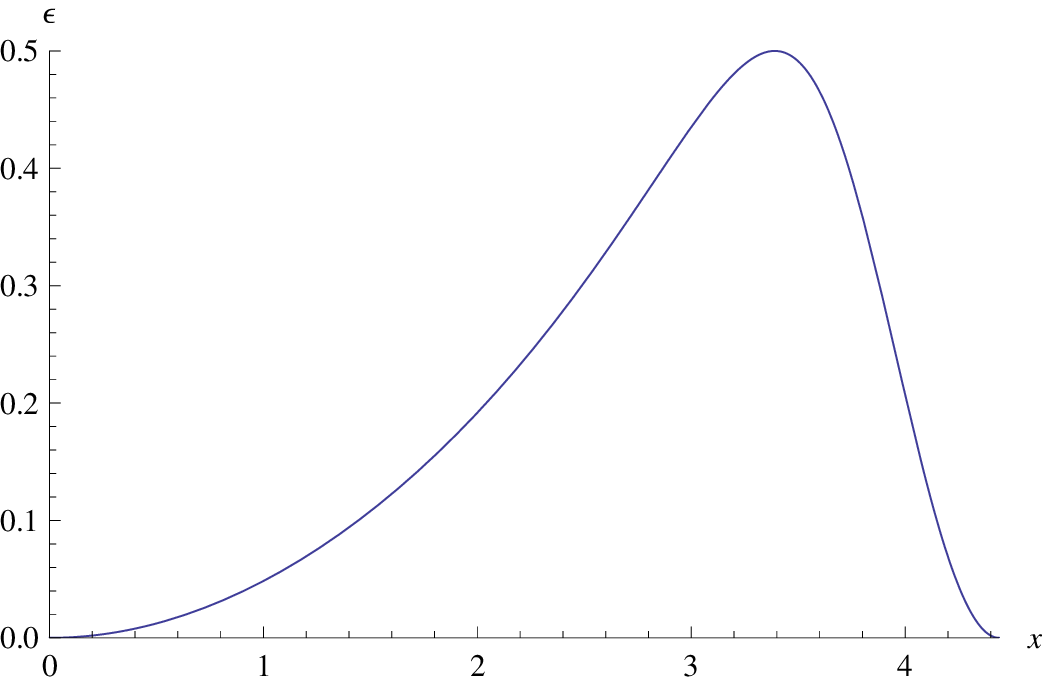}} \\
  \caption{Static solution for the new baby potential, with the coupling constant values $\mu = \gamma =\lambda = \rho = n = 1$.}
  \label{New}
 \end{center}
\end{figure}

Solutions are again compactons, so the above discussion just repeats ($h_2$ is now  a factor of 2 smaller, i.e., exactly equal to (\ref{h-2}), because now $V\sim h$). The result of a numerical integration is shown in figure \ref{New}. The free parameters and the energy are
\begin{equation}
  a_0 = -0.555984,  \qquad x_0 = 4.439, \qquad E_0=5.62986 ,
 \end{equation}
where $a_0$ and $E_0$ again coincide with their BPS values. 
\subsubsection{The potential $V=h^2$} 
\begin{figure}[h]
 \begin{center}
  \subfloat[Function $h$ with its derivative.]{\includegraphics[width=0.45\textwidth]{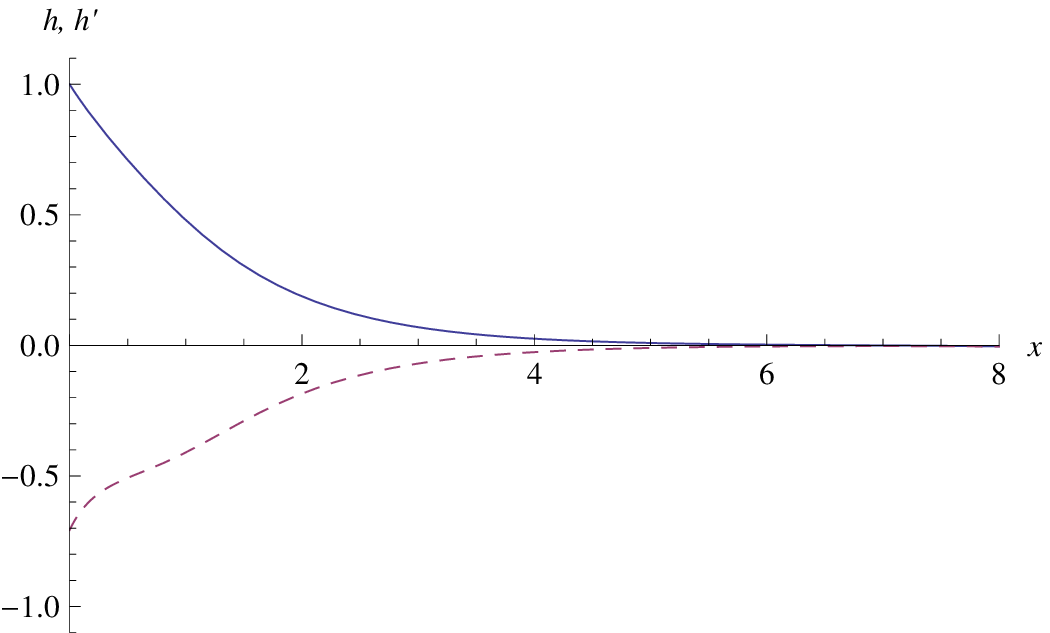}} \quad
  \subfloat[Function $a$ and its derivative.]{\includegraphics[width=0.45\textwidth]{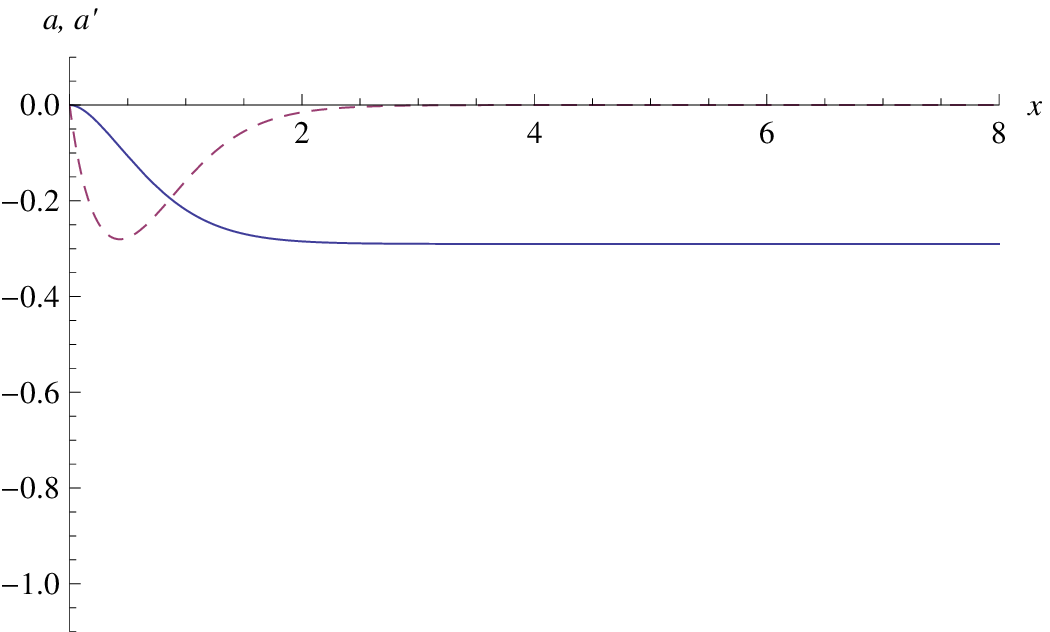}} \\
  \subfloat[Function $b$ and its derivative.]{\includegraphics[width=0.45\textwidth]{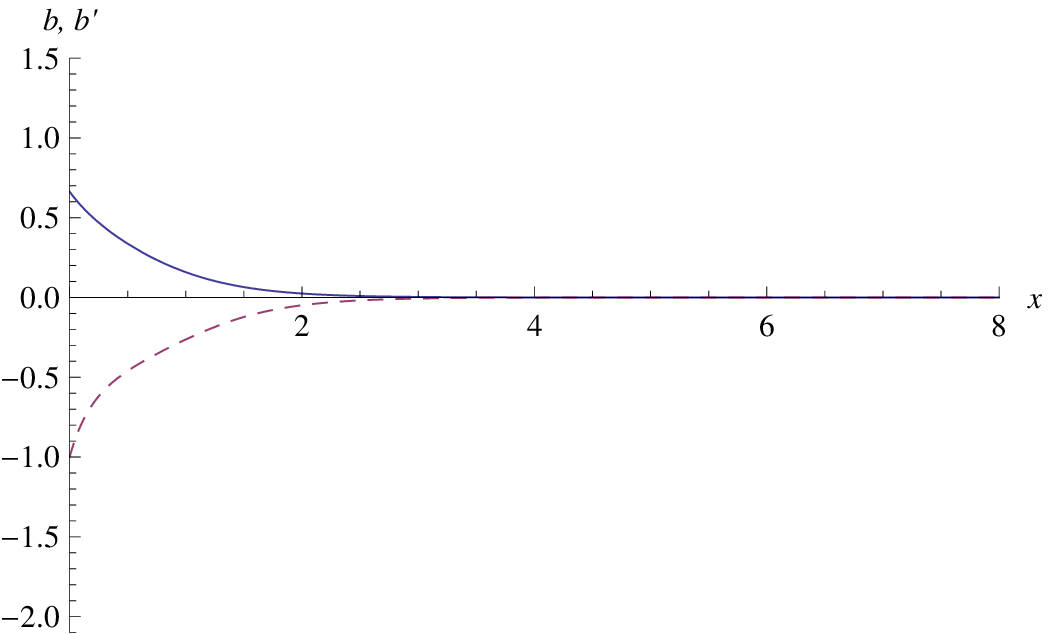}} \quad
  \subfloat[Energy density.]{\includegraphics[width=0.45\textwidth]{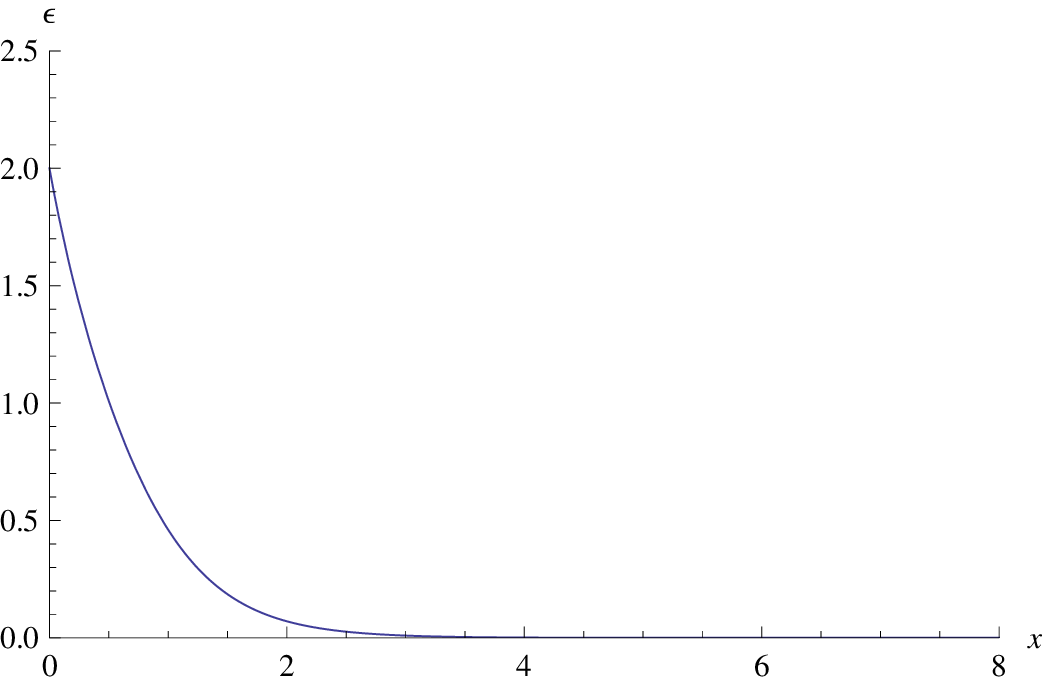}} \\
  \caption{Static solution for the potential $V=h^2$, with the coupling constant values $\mu = \gamma =\lambda = \rho = n = 1$.}
  \label{Vh2}
 \end{center}
\end{figure}

Here, the Skyrme field approaches its vacuum value exponentially, therefore we perform the shooting from the center. The expansion at the center gives
$$
h(x) \sim 1+h_1 x + \ldots
$$
$$
a(x) \sim \frac{2 \gamma b_0 h_1}{ \rho} x^2 + \ldots
$$
$$
b(x) \sim b_0 - \frac{2 \lambda^2 h_1^2 }{ \rho} x + \ldots
$$
where the free constants are $h_1$ and $b_0$.  For the numerical solution shown in figure \ref{Vh2} these constants and the energy take the values
$$
 h_1 = - 0.7071  , \quad 
 b_0 = 0.663114, \quad E_0 = 8.333
$$
and the energy is, again, equal to its BPS value. So, the explicit numerical calculations completely confirm our analytical results.

%%%%%%%%%%%%%%%%%%%%%%%%%%%%%%%%%%%%%%%%%
\section{Conclusions}
%%%%%%%%%%%%%%%%%%%%%%%%%%%%%%%%%%%%%%%
In this paper we investigated in detail some Lifshitz field theories, characterized by an anisotropic scaling between space and time, with dynamical critical exponent $z=d$, where $d$ is the dimension of space. We explicitly considered the case $d=2$ for reasons of simplicity. More concretely, we studied Lifshitz field theories which have a large (infinite-dimensional) group of base space symmetries, namely the Special Diffeomorphisms on $d$-dimensional space. One reason to consider these theories is that the SDiff group forms a relevant (infinite-dimensional) subgroup of the symmetry group of Horava--Lifshitz gravity \cite{G}, so these models may hopefully shed some light on non-perturbative properties of the latter theory. Another reason is that the presence of such a large group of symmetries typically allows to find exact solutions of the field equations, providing thereby analytical nonperturbative informations about these theories. For the class of theories considered, the relevant term in the lagrangian density (corresponding to a renormalizable interaction w.r.t. Lifshitz scaling) could be interpreted as the square of a topological charge density, so that topological soliton solutions can be expected. Concretely, we investigated theories where the fields had the topology of skyrmions, but, as said, a generalization of our results to the cases of vortices or monopoles should pose no problem. We found that these theories have a BPS bound and topological solitons, i.e., skyrmions, saturating the bound. The static sector of the theories is, in fact, equivalent to the static sector of the so-called BPS Skyrme models \cite{BPS-Sk}- \cite{Sp1}, where BPS skyrmions are known to exist. Time-dependent (e.g. Q-ball) solutions, on the other hand, behave qualitatively different in the Lorentz-invariant BPS Skyrme models and in the Lifshitz field theories considered here. Finally, we studied the case where the Lifshitz field theory is coupled to an abelian gauge field, where both scaling symmetry and renormalizability required a Chern--Simons  instead of a Maxwell gauge theory. We found that the resulting theory, again, has both a BPS bound and solutions saturating this bound where, as in the case of the gauged BPS baby Skyrme model \cite{gaugedBPSbaby}, the derivation of the BPS bound required the introduction of a superpotential, as in supergravity coupled to a scalar field \cite{susy1}, \cite{f susy}. 

The strongest motivation for the recent interest in Lifshitz field theories is probably related to the possiblitiy to find a renormalizable and unitary quantum theory of gravity, as proposed in Horava-Lifshitz gravity \cite{G}. But if an anisotropic scaling is assumed for gravity in the UV region, then most likely the universal character of gravity makes it necessary to extend this anisotropic scaling also to the remaining fields and interactions. This then   
requires a detailed study of the perturbative and nonperturbative aspects of the resulting Lifshitz field theories, both for a check of the consistency of the whole construction and in order to understand whether and to which degree Lifshitz field theories may give rise to new physical phenomena not present in the more standard theories with invariance under Lorentz transformations. The results of the present paper are part of this ongoing investigation.

The are several obvious directions in which the present work can be continued. First of all, as the Lifshitz BPS baby model has a standard time evolution, one may try to analyze the dynamical properties of the model. Since the model defines a well-defined Cauchy problem, one may, e.g.,  integrate the time evolution using some numerical methods. For example, one may be interested in scattering processes of the (exactly known) solitons or Q-balls. Such an analysis becomes even more relevant taking into account the infinitely large symmetry group of the model. Indeed, we have infinitely many conservation laws which may strongly influence the solitonic collisions. Hence, one may hope that a link between the standard integrability (e.g., manifested by a lack of radiation during scattering) and the generalized integrability in more than 1+1 dimensions can be found. The previously known examples of models integrable in the generalized sense do not have (globally) well-possessed Cauchy problems (Aratyn-Ferreira-Zimerman model \cite{AFZ}, BPS baby and BPS Skyrme) or are defined as constrained systems (integrable submodels) by some additional derivative dependent constraints. Therefore, it is very difficult to investigate dynamical properties of these theories, either analytically or numerically. As a consequence, it has not been verified yet whether the scattering of the corresponding solitons occurs in the same way as in the usual integrable systems (no radiations, the same particles in the initial and final states etc.). 

Another straightforward generalization is the construction and analysis of the analogous Lifshitz Skyrme model in (3+1) dimensions. 

\vspace*{0.3cm}

{\centerline {\bf Acknowledgement}}

\vspace*{0.2cm}

The authors acknowledge financial support from the Ministry of Education, Culture and Sports, Spain (grant FPA2008-01177), the Xunta de Galicia (grant INCITE09.296.035PR and Conselleria de Educacion), the Spanish Consolider-Ingenio 2010 Programme CPAN (CSD2007-00042), and FEDER. CN thanks the Spanish Ministery of Education, Culture and Sports for financial support (grant FPU AP2010-5772). Further, AW was supported by polish NCN grant 2011/01/B/ST2/00464.

\end{document}